\documentclass[prb,twocolumn,nopacs,superscriptaddress,floatfix,longbibliography]{revtex4-2}

\usepackage{epsfig}
\usepackage[centertags]{amsmath}
\usepackage{color,hyperref}
\usepackage{upgreek}
\usepackage[normalem]{ulem}

\definecolor{grey}{rgb}{.65,.65,.65}

\newcommand{\K}{\ensuremath{\,\mbox{K}}}

\newcommand{\Tc}{\ensuremath{T_{\rm C}}}
\newcommand{\Tn}{\ensuremath{T_{\rm N}}}

\begin{document}

\title{Multiferroic quantum criticality in (Eu,Ba,Sr)TiO$_3$ solid solution}

\author{Dalibor Rep\v{c}ek}
\thanks{Author to whom correspondence should be addressed}
\email[e-mail: ]{repcek@fzu.cz}
\affiliation{Institute of Physics, Czech Academy of Sciences, Na Slovance~2, 182 00 Prague~8, Czech Republic}
\affiliation{Faculty of Nuclear Sciences and Physical Engineering, Czech Technical University in Prague, B\v{r}ehová 7, 115 19 Prague 1, Czech Republic}
\author{Petr Proschek}
\affiliation{Faculty of Mathematics and Physics, Charles University, Ke Karlovu~5, 121 16 Prague~2, Czech Republic}
\author{Maxim Savinov}
\affiliation{Institute of Physics, Czech Academy of Sciences, Na Slovance~2, 182 00 Prague~8, Czech Republic}
\author{Martin Kachl\'{i}k}
\affiliation{CEITEC-Central European Institute of Technology, Brno University of Technology, Purky\v{n}ova 123, 612 00 Brno, Czech Republic}
\author{Ji\v{r}\'{i} Posp\'{i}\v{s}il}
\affiliation{Faculty of Mathematics and Physics, Charles University, Ke Karlovu~5, 121 16 Prague~2, Czech Republic}
\author{Jan Drahokoupil}
\affiliation{Institute of Physics, Czech Academy of Sciences, Na Slovance~2, 182 00 Prague~8, Czech Republic}
\author{Petr Dole\v{z}al}
\affiliation{Faculty of Mathematics and Physics, Charles University, Ke Karlovu~5, 121 16 Prague~2, Czech Republic}
\author{Jan Prokle\v{s}ka}
\affiliation{Faculty of Mathematics and Physics, Charles University, Ke Karlovu~5, 121 16 Prague~2, Czech Republic}
\author{Stanislav Kamba}
\email[e-mail: ]{kamba@fzu.cz}
\affiliation{Institute of Physics, Czech Academy of Sciences, Na Slovance~2, 182 00 Prague~8, Czech Republic}

\begin{abstract}
Based on the earlier published theory (\textit{Nature Mat}. \textbf{18}, 223--228 (2019)), a comprehensive experimental investigation of multiferroic quantum critical behavior of (Eu,Ba,Sr)TiO$_3$ polycrystalline and single crystal samples was performed. Presence of the displacive ferroelectric quantum criticality is revealed through non-classical ($T^2$) temperature scaling of inverse dielectric susceptibility up to 60\,K. With increasing hydrostatic pressure, this ferroelectric quantum criticality is gradually suppressed. Inverse magnetic susceptibility follows classical Curie-Weiss law down to $4\K$, but quantum fluctuations belonging to an antiferromagnetic phase transition ($\Tn < 0.8\K$) change its scaling below $3\K$ to $T^{(1.7\pm 0.1)}$ and $T^{(2.1\pm 0.2)}$ for samples containing 29\,\% and 25\,\% of Eu$^{2+}$ ions, respectively. Experimental indications of the coexisting ferroelectric and antiferromagnetic, i.e. multiferroic, quantum fluctuations and qualitative explanation why they could be seen only in the immediate proximity of $\Tn$ is given.

\end{abstract}
\date{\today}

\maketitle

\section{Introduction}

In recent years, considerable scientific effort was devoted to study of quantum phase transitions\cite{Sachdev11} (QPT), i.e. those which are triggered by quantum instead of thermal fluctuations characteristic for the classical phase transitions. It turns out that the quantum critical fluctuations may help to form new effective electron-electron interactions and thus support emergence of new phases such as unconventional superconductivity.\cite{Gegenwart08,Mathur98,Saxena00,Ran19} The quantum critical point (QCP) of the QPT needs to be located close to the temperature of absolute zero, where quantum fluctuations are dominant. Under these conditions, interplay of quantum and thermal fluctuations has an impact on finite-temperature behavior of the system such as a non-classical scaling of correlation functions.\cite{Lake05}

Dielectric materials are considered to be text-book systems for the description of ferroelectric (FE) QPT, since in conducting materials, the lattice fluctuations may be obscured by itinerant electrons.\cite{Rowley14} Quantum criticality theory applied to the FE QPT\cite{Roussev03} predicts unusual scaling of dielectric susceptibility ($\chi_{\boldsymbol{\upphi}}$) with respect to temperature. In three dimensional system, its reciprocal value should follow quadratic ($\chi_{\boldsymbol{\upphi}}^{-1}\propto T^{2}$) trend.\cite{Rowley14} This was indeed experimentally confirmed in two quantum paraelectric materials -- KTaO$_3$ and SrTiO$_3$. In the latter, $\chi_{\boldsymbol{\upphi}}^{-1}\propto T^{2}$ is satisfied up to $\approx$\,50$\K$, which demonstrates that the quantum critical fluctuations are relevant even tens of Kelvins above the QCP and may be even responsible for the emergence of superconductivity in SrTiO$_3$.\cite{Edge15}

Since there exist different materials exhibiting FE or one of the magnetic (ferromagnetic -- FM, antiferromagnetic -- AFM, etc.) QPT,\cite{Rowley10} it is a natural step further to try and combine these two QPTs in one physical system and create thus a multiferroic quantum criticality (MFQC).\cite{Narayan19,She10,Morice17} In order to do so, we may tune one or more physical parameters (pressure, substitution/doping, etc.) in a way that both critical points occur at 0$\K$. Similarly as for KTaO$_3$ and SrTiO$_3$, a potential existence of MFQC should be experimentally detectable e.g. via the non-classical temperature scaling of dielectric ($\chi_{\boldsymbol{\upphi}}$) and magnetic ($\chi_{\boldsymbol{\uppsi}}$) susceptibility of the MFQC material. While the FE QC causes the above mentioned $\chi_{\boldsymbol{\upphi}}^{-1}\propto T^{2}$ scaling, the quantum critical fluctuation of the magnetic order would lead to $\chi_{\boldsymbol{\uppsi}}^{-1}\propto T^{\frac{4}{3}}$ or $\chi_{\boldsymbol{\uppsi}}^{-1}\propto T^{\frac{3}{2}}$ in case of FM QCP or AFM QCP, respectively. \cite{Narayan19,Chandra17} Moreover, when the MFQC is present in the studied system, the QC spin and lattice excitations may even interact with each other.\cite{Flavian23} The mutual interaction of those two involved QC fields may effectively lead to a change of the original scaling of $\chi_{\boldsymbol{\upphi}}^{-1} (T)$ or $\chi_{\boldsymbol{\uppsi}}^{-1} (T)$. According to a type of the interaction (biquadratic in electric and magnetic order parameters, gradient - e.g. inverse Dzyaloshinskii-Moriya, etc.), one would expect for the FE+AFM MFQC possible crossover from $\chi_{\boldsymbol{\upphi}}^{-1}\propto T^{2}$ to $\chi_{\boldsymbol{\upphi}}^{-1}\propto T^{\frac{3}{2}}$ or other subdominant (higher-order) corrections to $\chi_{\boldsymbol{\upphi}}^{-1} (T)$ and $\chi_{\boldsymbol{\uppsi}}^{-1} (T)$ yielding $\chi_{\boldsymbol{\upphi}}^{-1}\propto T^{\frac{5}{2}}$ or $\chi_{\boldsymbol{\upphi}}^{-1}\propto T^{4}$ and $\chi_{\boldsymbol{\uppsi}}^{-1}\propto T^{2}$.\cite{Narayan19}

The (Eu,Ba,Sr)TiO$_3$ solid solution was earlier proposed as a promising candidate for achieving the MFQC at normal pressure using the mutual concentration of Eu$^{2+}$, Ba$^{2+}$ and Sr$^{2+}$ ions as the tuning physical parameter.\cite{Narayan19} The exact chemical composition, where the MFQC should exhibit the strongest interaction of the FE and AFM quantum fluctuations was predicted to be Eu$_{0.3}$Ba$_{0.1}$Sr$_{0.6}$TiO$_3$.

However, the MFQC has never been experimentally studied in this (or other) system before and therefore we focus on it in this paper. Let us first briefly describe key properties of each component of the compound to understand the origin of the predicted MFQC here. 

\section{Material description}

The (Eu,Ba,Sr)TiO$_3$ (EBSTO) solid solution proposed for the MFQC\cite{Narayan19} is largely based on SrTiO$_3$, which makes about 60\,\% of the whole compound. SrTiO$_3$ is a good starting point for reaching the MFQC, since it already exhibits soft-mode-driven FE QC itself.\cite{Rowley14} Diamagnetic SrTiO$_3$ is a well-known perovskite-like oxide,\cite{Brous53} meaning that at room temperature it crystallizes in cubic $Pm\overline{3}m$ structure. Then at 105$\K$ undergoes a structural phase transition (PT) to a non-polar tetragonal $I4/mcm$ phase due to a unit cell doubling resulting from rotation of the oxygen octahedra.\cite{Muller68,Fleury68} Its dielectric permittivity is decribed by Curie-Weiss law with theoretical critical temperature of 35\,K,\cite{Weaver59} but the quantum fluctuations prevent the anticipated FE phase transition and the permittivity exhibits gradual saturation on the value of 25 000 near 10 K.\cite{Muller79,Weaver59}

EuTiO$_3$ is closely related to SrTiO$_3$, since it is also the perovskite-like oxide with very similar lattice parameter defining its cubic $Pm\overline{3}m$ structure.\cite{Brous53} The antiferrodistortive PT to tetragonal $I4/mcm$ structure takes place at somewhat higher temperature of 282$\K$.\cite{Bussmann-Holder11,Goian12a,Kohler12} Similarly to SrTiO$_3$, the ferroelectricity is not present at low temperatures in EuTiO$_3$. It is an incipient ferroelectric, since even the high-temperature, Curie-Weiss behavior of dielectric susceptibility shows negative critical temperature ($-185\K$)\cite{Goian09} suggesting that EuTiO$_3$ would not have reached ferroelectric phase even without quantum fluctuations at low temperatures. Note that the negative Curie temperature in EuTiO$_3$ has no connection to antiferroelectricity (as could be expected from direct parallel to antiferromagnetism). Antiferroelectrics usually exhibit positive Curie temperature,\cite{Roleder96} while negative critical temperature is typical for incipient ferroelectrics. Importantly, EuTiO$_3$ brings to the resulting EBSTO system magnetism via Eu$^{2+}$ ions possessing a large magnetic dipole moment of seven Bohr magnetons. Pure EuTiO$_3$ exhibits $G$-type antiferromagnetism below the N\'{e}el temperature of $\Tn=5.3\K$ with competing AFM-like nearest-neighbor and FM-like next-nearest-neighbor interactions.\cite{McGuire66,Scagnoli12,Ryan13}

BaTiO$_3$ is a well-known, but still frequently studied text-book ferroelectric material with Curie temperature of almost $\Tc\approx 400 \K$. It undergoes two more ferroelectric PTs below $\Tc$ resulting in a trigonal polar phase below $\approx$\,190$\K$.\cite{Hippel50} Since the solid solution of EuTiO$_3$ and SrTiO$_3$ is not so close to the FE QCP as the pure SrTiO$_3$ is, the BaTiO$_3$ is used in the eventual EBSTO system to bring it closer to the FE QCP. Then it should be theoretically possible to combine SrTiO$_3$, EuTiO$_3$ and BaTiO$_3$ in a way that the EBSTO system will retain the FE QC of the pure SrTiO$_3$ and simultaneously exhibit the magnetic (AFM) QC coming from AFM critical point of pure EuTiO$_3$ tuned to 0$\K$.\cite{Narayan19}

Various EuTiO$_3$, SrTiO$_3$, BaTiO$_3$ solid solutions have been previously investigated in terms of a spin-phonon coupling, magnetodielectric effect, or even multiferroicity.\cite{Katsufuji01,Rushchanskii10,Goian13,Bussmann-Holder15} Nevertheless, quantum critical phenomena have never been the subject of experimental research on this system.

\section{Sample preparation}

As starting materials, pure powders of SrCO$_3$, Eu$_2$O$_3$, BaCO$_3$, TiO$_2$ (anatas) were used for preparation of the desired stoichiometric mixtures. The mixtures were homogenized by the low-energy planetary ball milling lasting for 48 hours. The zirconia balls as milling elements and ethanol as a medium were used. The homogenized mixtures were dried at 50\,$^\circ$C for 5 days and subsequently at 80\,$^\circ$C for 2 days. The dried compacted agglomerates were then cautiously crushed to powder again. The solid state reaction of precursors was performed via annealing in reducing conditions of hydrogen atmosphere (99.9\,\% H$_2$). The powder products of annealing were formed into discs (16\,mm in diameter; weight $\approx$ 1\,g) by uniaxial pressure of 10\,MPa followed by 1000\,MPa of cold isostatic pressure (CIP). The sintering process of the green bodies shaped by CIP was performed in the conventional furnace with tungsten heating elements at 1400\,$^\circ$C for 2 hours in either argon (950\,mbar) or vacuum ($\approx$ 10$^{-5}$\,mbar) environment. All the polycrystalline (ceramic) samples with the same stoichiometry subsequently exhibited very similar physical properties, regardless of the sintering environment used. Mass densities of the ceramics after sintering as well as the contents of open resp. closed porosity were measured by the Archimedes method (EN 623-2). Relative density of all ceramics reached about 95\,\%.

Eu$_{0.29}$Ba$_{0.08}$Sr$_{0.63}$TiO$_3$ single crystal was prepared by the floating zone method (Fig. S1 in Supplemental material\cite{Suppl}) which is very effective for the growth of large high-quality single crystals of various materials.\cite{Smilauerova14,Zajic23} An advanced laser diode optical furnace with 5 lasers (model  FZ-LD-5-200W-VPO-PC-EG, Crystal Systems Corp., Japan) was used. In the first step, a polycrystalline (ceramic) material was synthesized as described in previous paragraph. Then, a precursor in the form of a 50\,mm long sintered rod with diameter of 4\,mm was prepared. The quartz chamber of the optical furnace was evacuated by a turbomolecular pump to $10^{-6}$\,mbar before the crystal growth process. The whole growing process was performed in 6N argon flow. The pulling rate was slow only 1\,mm/h with rotation -3/30\,rpm (feed rod and single crystal rotate in opposite direction with 3 and 30 rpm, respectively). A single crystal of the cylindrical shape with length of $\approx 30$\,mm and varying diameter of $3-5$\,mm was obtained (Fig. S2 in Supplemental material\cite{Suppl}). The quality of the single crystal was confirmed by Laue diffraction showing sharp reflections in the pattern (Fig. S3 in Supplemental material\cite{Suppl}). The Laue diffraction revealed a 35$^{\circ}$ tilt of the cubic crystallographic axis from the main axis of the cylindrical single crystal sample.

The actual composition of our samples was in good agreement with the intended chemical composition. This was checked using Wavelength Dispersive Spectroscopy (WDS) on many sites inside the EBSTO single crystal. Here, the stoichiometry of the chemical reaction was balanced in a way to obtain the desired Eu$_{0.3}$Ba$_{0.1}$Sr$_{0.6}$TiO$_3$ solid solution. The WDS revealed the actual composition to be Eu$_{0.29}$Ba$_{0.08}$Sr$_{0.63}$TiO$_3$, which also proves reliability of the preparation process of ceramics, whose composition was determined from the stoichiometry of the precursor mixture.

\section{Experimental details}

The low temperature X-ray diffraction was measured by powder $\theta - \theta$ diffractometer Siemens D500 in Bragg-Brentano geometry using the Cu$_{\mathrm{K}_{\alpha 1,2}}$ radiation. He closed-cycle system was used for cooling. The sample was placed in the low-temperature cryostat (ColdEdge) on the sapphire cube. Sample chamber was filled with He atmosphere to ensure sufficient thermal contact with the cold finger. Temperature stability was better than $0.1\K$. The diffracted intensity was measured by a linear position sensitive detector Mythen 1K, the measured data were processed using the procedures described elsewhere,\cite{Kriegner15} and finally lattice parameters were determined using the Le Bail analysis. Temperature dependence of lattice parameters of the Eu$_{0.3}$Ba$_{0.1}$Sr$_{0.6}$TiO$_3$ ceramics and comparison of the Bragg peak profile at room and helium temperature confirmed stability of the cubic $Pm\overline{3}m$ structure at least down to $5\K$ (Figs. S4, S5 in Supplemental material\cite{Suppl}). It is quite surprising, since EuTiO$_3$, SrTiO$_3$, and their Eu$_x$Sr$_{1-x}$TiO$_3$ solid solution exhibit antiferrodistortive phase transition at relatively high temperatures between $280$ and $105\K$.\cite{Bussmann-Holder15} This shows that $10$\,$\%$ of Ba effectively stabilizes the cubic phase in the (Eu,Ba,Sr)TiO$_3$.

The dielectric properties were measured by Novocontrol Alpha-AN high performance impedance analyzer and ANDEEN-HAGERLING Ultra-precision Capacitance Bridge AH 2550A on PPMS device (Physical Properties Measurement System - Quantum Design) which was used for controlling external parameters - temperature (with $^3$He insert down to $0.4\K$) and magnetic field (up to 9\,T). For the application of pressure, the double-layered piston cylindrical pressure cell with a nominal pressure range of 3\,GPa was used.\cite{Fujiwara07} In all cases, manganin wire as a pressure gauge and Daphne 7373 as a pressure transmitting medium were used.\cite{Yokogawa07,Murata97,Stasko20} Electric field was always applied perpendicular to the sample plane.

Heat capacity measurements were performed using the heat capacity option in PPMS with $^3$He insert down to $0.4\K$ and up to 9\,T.

Magnetic properties were measured on PPMS device as well, using the vibrating sample magnetometer (VSM) option down to $2\K$ and up to 9\,T. Magnetization measurements below $2\K$ were performed via PPMS with $^3$He insert (temperature down to $0.4\K$) using Hall probe magnetometry (placing the sample on the high-sensitivity Hall probe and scaling the resulting data in an overlapping region with data observed by other methods, in our case via VSM). A custom-made detection coil used for measurements of AC magnetization could not be exactly calibrated and therefore its output values are relative. For the quantitative analysis, the AC magnetic susceptibility data were then normalized to the measured DC magnetic susceptibility, whose values are absolute. Since the aim of this work is to characterize temperature scaling of (the inverse) magnetic susceptibility of the cubic EBSTO system in the paramagnetic phase, i.e. in an isotropic environment, neither the orientation of external magnetic field, nor exact compensation of the demagnetizing field were of crucial importance here. However, magnetic field was always applied in the sample plane, in order to minimize effect of the demagnetizing field.

\section{Results and discussion}

\subsection{Ferroelectric Quantum criticality}

\begin{figure}[]
	\centering
	\includegraphics[width=87mm]{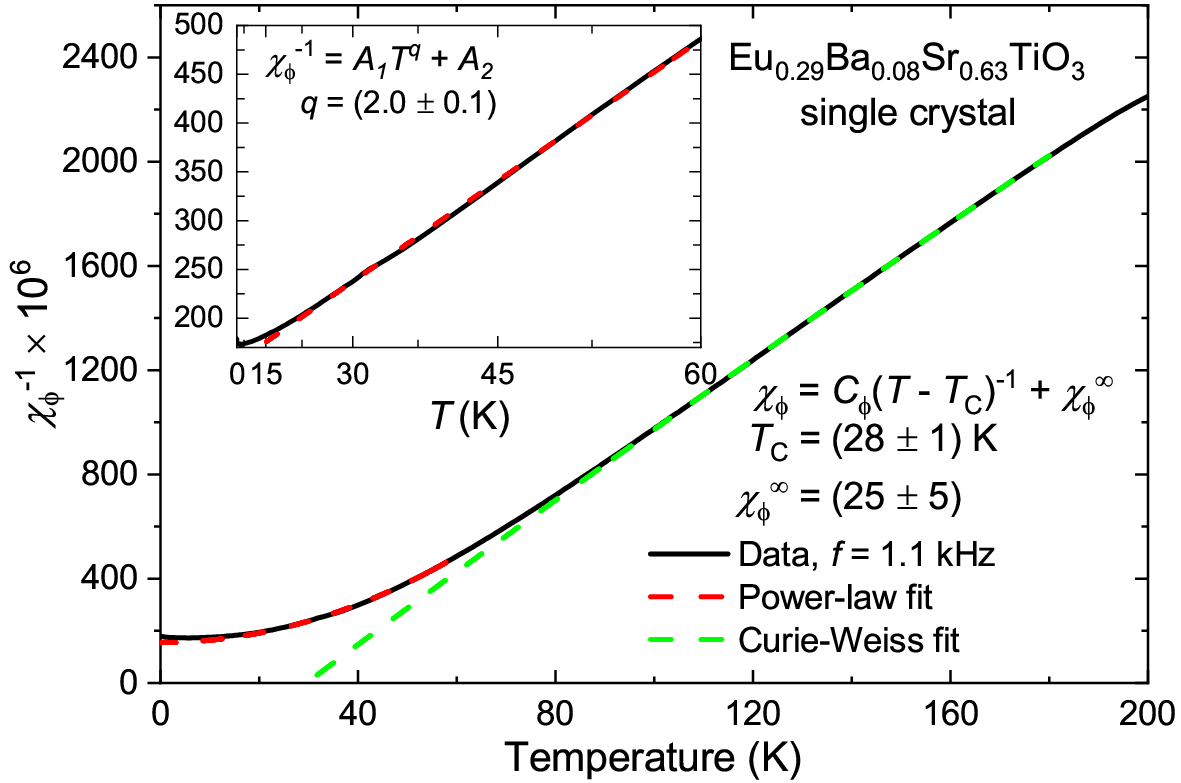}
	\caption{Temperature dependence of inverse dielectric susceptibility measured at 1\,kHz on the EBSTO single crystal without external magnetic field and its power-law and Curie-Weiss fit. Inset: The low-temperature range (displayed in quadratic temperature scale), where the FE QC takes place. Parameters of both low-temperature power-law and high-temperature C-W fits are $A_1=(0.092 \pm 0.003)\cdot 10^{-6}$,  $A_2=(155 \pm 1)\cdot 10^{-6}$, $q=(2.0 \pm 0.1)$, $C_{\boldsymbol{\upphi}} = (71 \pm 1)\cdot 10^{3}\K$, $T_{\mathrm{C}}=(28 \pm 1)\K$, and~$\chi_{\boldsymbol{\upphi}}^{\infty}=(25 \pm 5)$. The power-law fit was performed in the temperature range from 25 to 60\,K and its extrapolation depicted down to 0\,K.}
	\label{fig:inv_eps1_on_T_single_cryst}
\end{figure}

The low-frequency (kHz) dielectric susceptibility indeed exhibit a crossover from classical Curie-Weiss (C-W) critical behavior ($\chi_{\boldsymbol{\upphi}} (T) = \frac{C_{\boldsymbol{\upphi}}}{T-\Tc} + \chi_{\boldsymbol{\upphi}}^\infty$, where $C_{\boldsymbol{\upphi}}$ is Curie constant, $\Tc$ Curie temperature, and $\chi_{\boldsymbol{\upphi}}^\infty$ possible high-temperature shift) to the quantum critical behavior as expected. It is best seen from the plot of inverse dielectric susceptibility ($\chi_{\boldsymbol{\upphi}}^{-1} (T)$) of the EBSTO single crystal (Fig. \ref{fig:inv_eps1_on_T_single_cryst}). The classical (high-temperature) C-W region is characterized by the linear temperature dependence $\chi_{\boldsymbol{\upphi}}^{-1}\propto T$ above 80 K. The C-W fit performed from $80\K$ to $180\K$ gives hypothetical Curie temperature $\Tc = 28\K$, which is value only slightly lower than in pure SrTiO$_3$ ($35\K$).\cite{Weaver59} The EBSTO single crystal with composition Eu$_{0.29}$Ba$_{0.08}$Sr$_{0.63}$TiO$_3$ is therefore roughly as close to the QCP as pure SrTiO$_3$. This corresponds to the comparable region of the QC behavior of $\chi_{\boldsymbol{\upphi}} (T)$. In EBSTO it is observed up to $60\K$ and in SrTiO$_3$ up to 50\K.\cite{Rowley14} The predicted ideal content\cite{Narayan19} of 10\,\% of Ba$^{2+}$ ions could indeed result in even stronger FE quantum fluctuations and broader temperature region of the QC behavior, since the whole system would be tuned closer to the FE QCP. Interestingly, the QC behavior is in good agreement with $\chi_{\boldsymbol{\upphi}}^{-1}\propto T^{2}$ scaling predicted for the FE QC alone. No sign of a possible crossover to different QC regime due to the presence of magnetic QC is clearly visible, although small deviations from $\chi_{\boldsymbol{\upphi}}^{-1}\propto T^{2}$ may be hidden in the uncertainty of the fitting curve.

The necessity of describing the dielectric response at low temperatures by a power-law trend characterizing the FE QC behavior based on coupling between the different wavevector modes of the transverse-optical phonon branch\cite{Rowley14} is illustrated by the use of obviously inadequate Barrett formula\cite{Barrett52} (Fig. S6) for quantum paraelectrics. The fitting curve does not reproduce the data in the low-temperature region (nor in the range from $20$ to $40\K$), where the FE QC behavior takes place.

Confirmation of the FE QC behavior in EBSTO is also brought by the data measured on ceramic samples with chemical compositions Eu$_{0.3}$Ba$_{0.1}$Sr$_{0.6}$TiO$_3$ and Eu$_{0.25}$Ba$_{0.1}$Sr$_{0.65}$TiO$_3$ (see Figs. S7, S8 in Supplemental material\cite{Suppl}). Interestingly, Eu$_{0.25}$Ba$_{0.1}$Sr$_{0.65}$TiO$_3$ ceramics exhibits very similar hypothetical $\Tc = 33\K$ and also the FE QC behavior up to $60\K$ (Fig. S7). In Eu$_{0.3}$Ba$_{0.1}$Sr$_{0.6}$TiO$_3$ ceramics, the FE quantum fluctuation are weaker as the FE QC behavior may be observed in narrower temperature range (Fig. S8), which corresponds to smaller value of hypothetical $\Tc = 1\K$. This may be due to slight difference between stoichiometric compositions (mainly Ba$^{2+}$ content) of individual samples. Also other low-frequency dynamical mechanisms contributing to permittivity (see small, but frequency-dependent maximum in $\chi_{\boldsymbol{\upphi}}(T)$ above $40\K$ in Fig. S9 in Supplemental material\cite{Suppl}), which make the curve-trend analysis difficult, are more distinctive in ceramics compared to the single crystal. On the other hand, comparison of dielectric measurements of single crystal and ceramics indicate that the chemical disorder (which may be expected to be higher in ceramics due to grain boundaries) at the (Eu,Ba,Sr) position does not prevent the existence of the FE QC. This result is consistent with previous findings on SrTiO$_3$ ceramics.\cite{Rowley14} The chemical disorder is, of course, present even in the EBSTO single crystal. No superstructure was observed in X-ray diffraction.

\begin{figure}[]
	\centering
	\includegraphics[width=84mm]{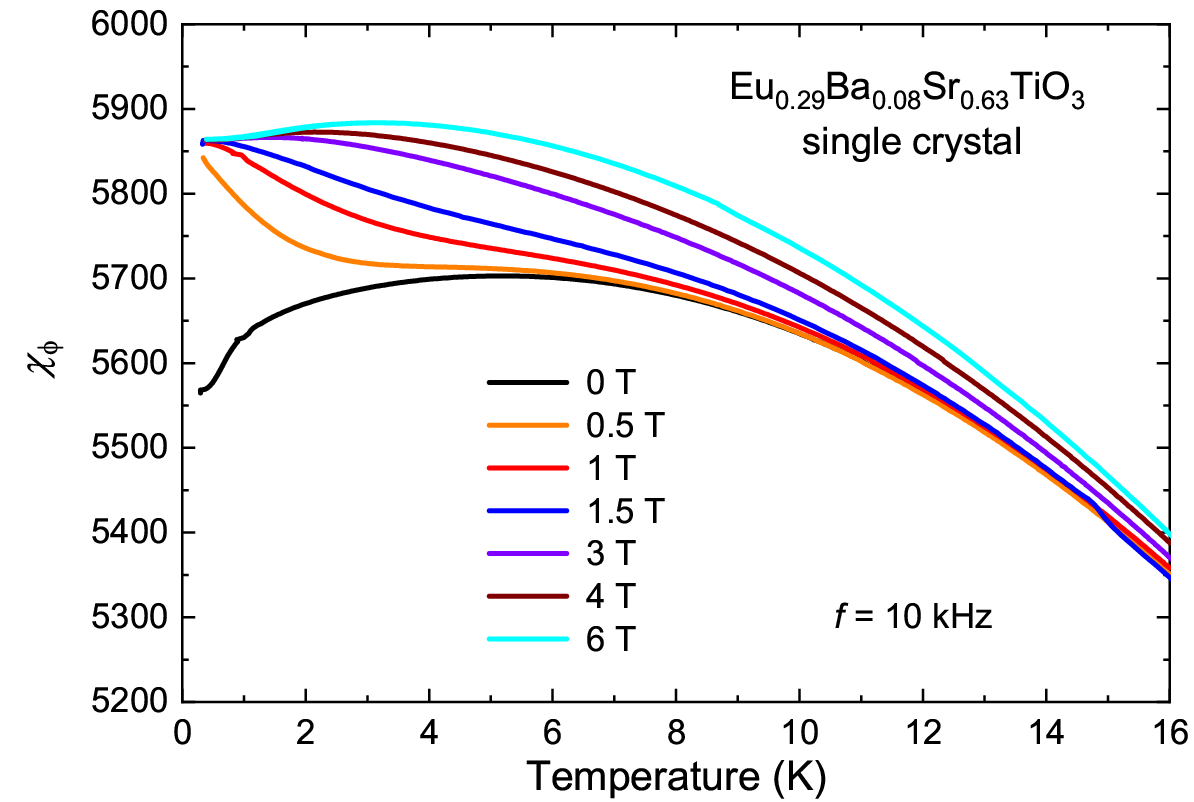}
	\caption{Temperature dependence of the 10\,kHz dielectric susceptibility of the EBSTO single crystal measured in various static external magnetic fields.}
	\label{fig:perm_v_mag_poli}
\end{figure}

\begin{figure}[]
	\centering
	\includegraphics[width=84mm]{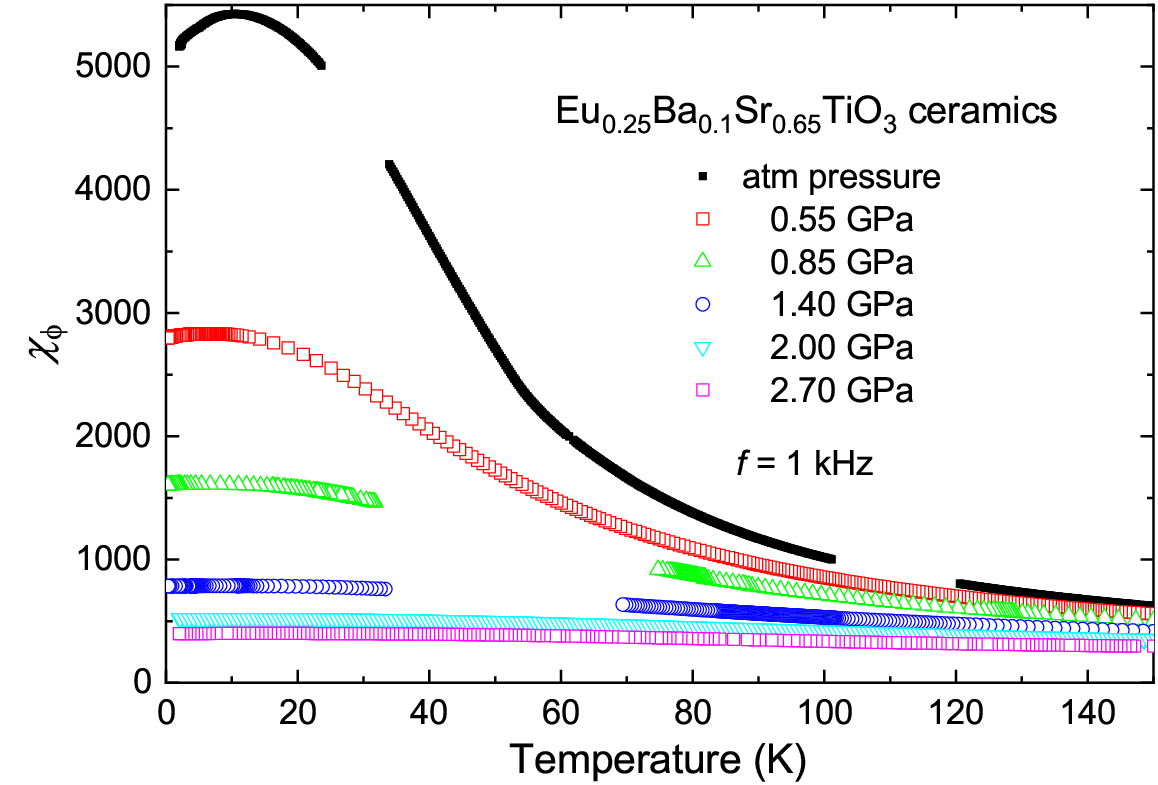}
	\caption{Temperature dependence of the 1\,kHz dielectric susceptibility measured at various hydrostatic pressures on the Eu$_{0.25}$Ba$_{0.1}$Sr$_{0.65}$TiO$_3$ ceramics.}
	\label{fig:eps1_on_T_various_pressures}
\end{figure}

Representing the measured data by $\chi_{\boldsymbol{\upphi}} (T)$ instead of $\chi_{\boldsymbol{\upphi}}^{-1} (T)$ (Figs. \ref{fig:perm_v_mag_poli}, \ref{fig:eps1_on_T_various_pressures}) highlights two features. First, the $\chi_{\boldsymbol{\upphi}} (T)$ reaches maximum value over 5000, which is less than 30\,\% of the value which is exhibited by the SrTiO$_3$,\cite{Muller79} but $15$ times higher than in EuTiO$_3$.\cite{Repcek20} Second, a shallow maximum appears at $\approx$\,10$\K$, which is not a sign of the FE ordering. Also pure SrTiO$_3$ features such maximum,\cite{Fischer85,Rowley14} but it is narrower, occurs at even lower temperatures ($\approx$\,2$\K$), and its possible explanation is a coupling of the soft optical phonon with the acoustic one.\cite{Rowley14} In some cases, the maximum in $\chi_{\boldsymbol{\upphi}}(T)$ was observed at higher temperatures than $\approx$\,2$\K$ and then it was attributed to an extrinsic effect resulting from heterogenity of the sample.\cite{Vendik97} However, Vendik's model\cite{Vendik97} of effective susceptibility of non-homogeneous samples is not satisfactory when applied to our data of EBSTO system. Here, the maximum is mainly due to a spin-phonon coupling inherited from pure EuTiO$_3$,\cite{Katsufuji01} which effectively reduces $\chi_{\boldsymbol{\upphi}}$ (and thus obscure influence of quantum fluctuations on $\chi_{\boldsymbol{\upphi}}(T)$ scaling) below and to some extent even several Kelvins above the Néel temperature $\Tn$ of the AFM PT.\cite{Repcek20} It is very well seen from the temperature dependence of $\chi_{\boldsymbol{\upphi}}$ measured in external magnetic field (Fig. \ref{fig:perm_v_mag_poli}). The spin-phonon coupling causes decrease of the zero-field dielectric susceptibility at low temperatures, while the magnetodielectric effect, which is based on this coupling, then increase the dielectric susceptibility in the same temperature range when the magnetic field is applied.\cite{Katsufuji01} The relative change of permittivity of the EBSTO single crystal with increasing magnetic field reaches $5$\,\% at the lowest temperature. It is similar value to the pure EuTiO$_3$ where a record high 7\,\% magnetodielectric effect was observed.\cite{Katsufuji01} However, even shallower maximum reappears in Fig. \ref{fig:perm_v_mag_poli} near $3\K$ in strong magnetic fields, where the magnetodielectric effect is saturated. This seems to be caused by the already known coupling of the optical and acoustic phonon branch,\cite{Rowley14} possibly combined with the effect of sample heterogenity.\cite{Vendik97} In any case, the existence of the shallow minimum in $\chi_{\boldsymbol{\upphi}}^{-1} (T)$ that is not directly related to the FE QC means that it is necessary to investigate the temperature dependence of $\chi_{\boldsymbol{\upphi}}^{-1} (T)$ far enough from this minimum, which in fact deviate the low-temperature $\chi_{\boldsymbol{\upphi}}^{-1} (T)$ dependence. The low-temperature limit for the observation of the FE QC in the EBSTO single crystal is then $\approx$ 20\,K. Nevertheless, this does not necessarily mean that the FE QC behavior is not present below this temperature limit. It may just be superimposed on the other phenomena (spin-phonon coupling, acoustic-optical phonon coupling, sample non-homogenity).

When hydrostatic pressure is applied on the Eu$_{0.25}$Ba$_{0.1}$Sr$_{0.65}$TiO$_3$ ceramics, it behaves in a qualitatively same way as pure SrTiO$_3$.\cite{Coak19} The magnitude of $\chi_{\boldsymbol{\upphi}} (T)$ gradually decreases with increasing pressure (Fig.~\ref{fig:eps1_on_T_various_pressures}) as a consequence of soft phonon hardening. It means that the FE QC behavior is gradually suppressed with increasing hydrostatic pressure. It demonstrates proximity of the Eu$_{0.25}$Ba$_{0.1}$Sr$_{0.65}$TiO$_3$ ceramic sample to the QCP. Analysis of the small and broad peak position in $\chi_{\boldsymbol{\upphi}} (T)$ is complicated because of the aforementioned spin-phonon coupling influencing its exact shape. Therefore, no simple dependence of the peak position is observed in contrast to SrTiO$_3$.\cite{Coak20}

\subsection{(Antiferro)magnetic quantum criticality}

\begin{figure}[]
	\centering
	\includegraphics[width=84mm]{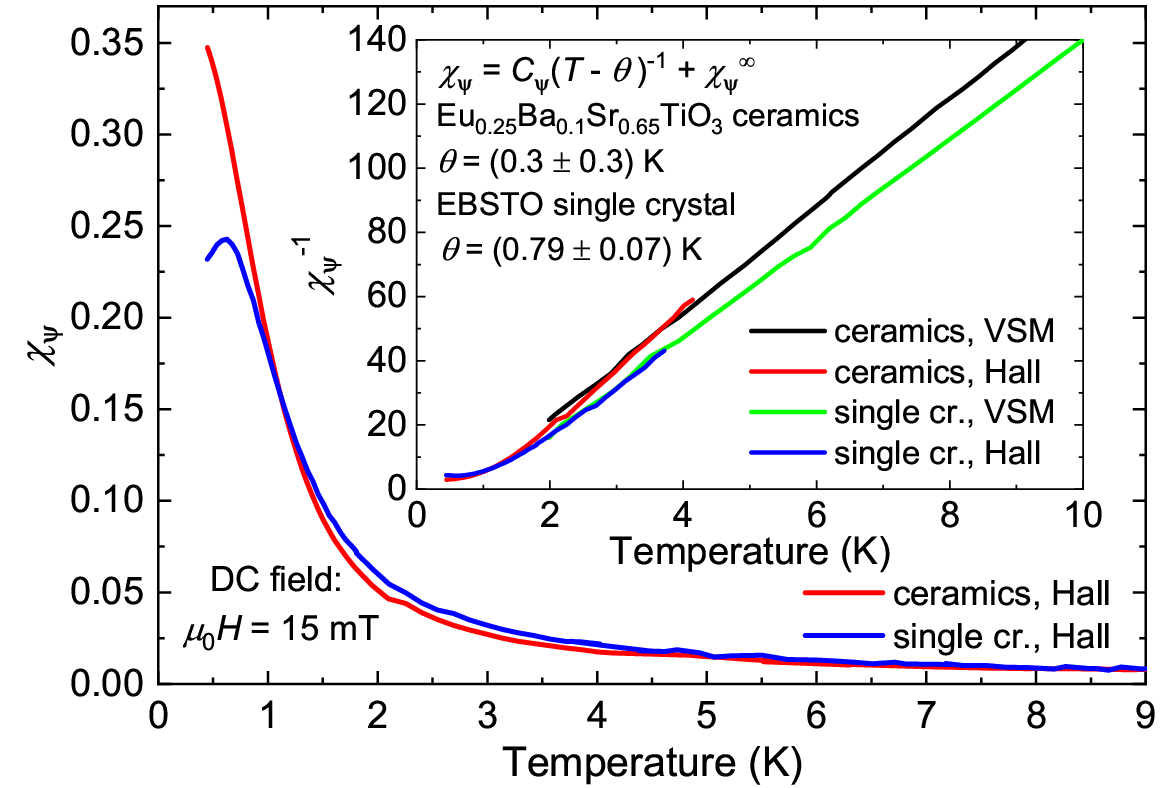}
	\caption{Temperature dependence of magnetic susceptibility (main graph) and its inverse (inset) measured in weak DC magnetic field of 15\,mT on the EBSTO single crystal and Eu$_{0.25}$Ba$_{0.1}$Sr$_{0.65}$TiO$_3$ ceramics. The data of $\chi_{\boldsymbol{\uppsi}}^{-1}$ follow the C-W dependence (for sake of clarity, fit not shown here) in temperature range $4-200\K$ with antiferromagnetic Curie temperature of $\theta = (0.79 \pm 0.07)\K$ for the single crystal and $\theta = (0.3 \pm 0.3)\K$ for the ceramics.\cite{explanation} Remaining fitting parameters converged to values $C_{\boldsymbol{\uppsi}} = (657 \pm 5)\cdot 10^{-4}\K$, $\chi_{\boldsymbol{\uppsi}}^{\infty}=(0 \pm 1)\cdot 10^{-6}$ and $C_{\boldsymbol{\uppsi}} = (610 \pm 10)\cdot 10^{-4}\K$, $\chi_{\boldsymbol{\uppsi}}^{\infty}=(8 \pm 5)\cdot 10^{-6}$, respectively.}
	\label{fig:magn_susc_and_inv_magn_susc_on_T}
\end{figure}

Analysis of the magnetic QC behavior rely on measurement of temperature dependent magnetization ($M$) and its consequent conversion to magnetic susceptibility ($\chi_{\boldsymbol{\uppsi}}=M/H$). Temperature dependence of $\chi_{\boldsymbol{\uppsi}}$ measured on the EBSTO single crystal in weak static magnetic field of 15\,mT (Fig. \ref{fig:magneticka_susceptibilita_single_crystal}) reveals a peak at 0.6\K. Based on the evidences given below, we attribute it to the AFM PT (due to Eu$^{2+}$ ions of EuTiO$_3$). First indication is shift of the peak towards low temperature limit when the strength of the external magnetic field is increased. It is also confirmed from the data of AC magnetic response (Fig. S10 in Supplemental material\cite{Suppl}), where, strictly speaking, differential magnetic susceptibility is measured. The peak is located already on the low-temperature edge of our measuring range in weak magnetic field of 50 mT and then disappears below it in stronger field. This is consistent with a suppression of the AFM ordering. Second evidence speaking in favor of magnetic PT is a shift of the peak to lower temperatures with decreasing content of Eu$^{2+}$ ions (see the comparison of magnetic susceptibility of EBSTO single crystal with 29\,\% of Eu$^{2+}$ ions and ceramics with 25\,\% of Eu$^{2+}$ ions in Fig. \ref{fig:magn_susc_and_inv_magn_susc_on_T}). These results are also consistent with data of magnetic moment measured on ceramics with 30\,\% of Eu$^{2+}$ ions (Fig. S11 in Supplemental material\cite{Suppl}), where the peak is located at 0.7$\K$ in magnetic field of 15 mT. Stable position of the peak in AC magnetic susceptibility of the single crystal (Fig. S12 in Supplemental material\cite{Suppl}) at 0.7$\K$ (in zero DC magnetic field) for various frequencies of the AC magnetic field then implies that this PT is not of a magnetic-glass type.

In order to bring another piece of information and check $\Tn$, the measurement of heat capacity was performed on both ceramics (Fig. \ref{fig:tepelna_kapacita_stara_keramika} and Figs. S14, S15 in Supplemental material\cite{Suppl}) and also single crystal (Fig. S16 in Supplemental material). The pure magnetic contribution to heat capacity ($C_{\mathrm{mag}}$ in Fig. \ref{fig:tepelna_kapacita_stara_keramika}), which is further divided by temperature in Fig. S14, was obtained by subtracting a phonon background $\gamma T^3 + \delta T^5$ from the total measured heat capacity.\cite{Petrovic13} Indeed, the zero-magnetic-field data reveal a peak at low temperatures in all cases (except for the single crystal, where we do not have reliable data below $0.8\K$). Position of the peak ($0.7-0.8\K$) for ceramics with 30\,\% of Eu$^{2+}$ ions (Fig. \ref{fig:tepelna_kapacita_stara_keramika}) coincide with the peak in magnetic moment measured in weak DC magnetic field of 15 mT (Fig. S11 in Supplemental material\cite{Suppl}) and may be thus attributed to the AFM PT. The same result applies to the ceramics with 25\,\% of Eu$^{2+}$ ions, where the peak in heat capacity lies at $0.5-0.6\K$ (Fig. S15 in Supplemental material\cite{Suppl}), i.e. lower than in the compound with higher concentration of Eu$^{2+}$ ions. The peak position for the single crystal and related AFM transition lies below $0.8\K$, a technical limit of the experiment for this sample.

\begin{figure}[]
	\centering
	\includegraphics[width=84mm]{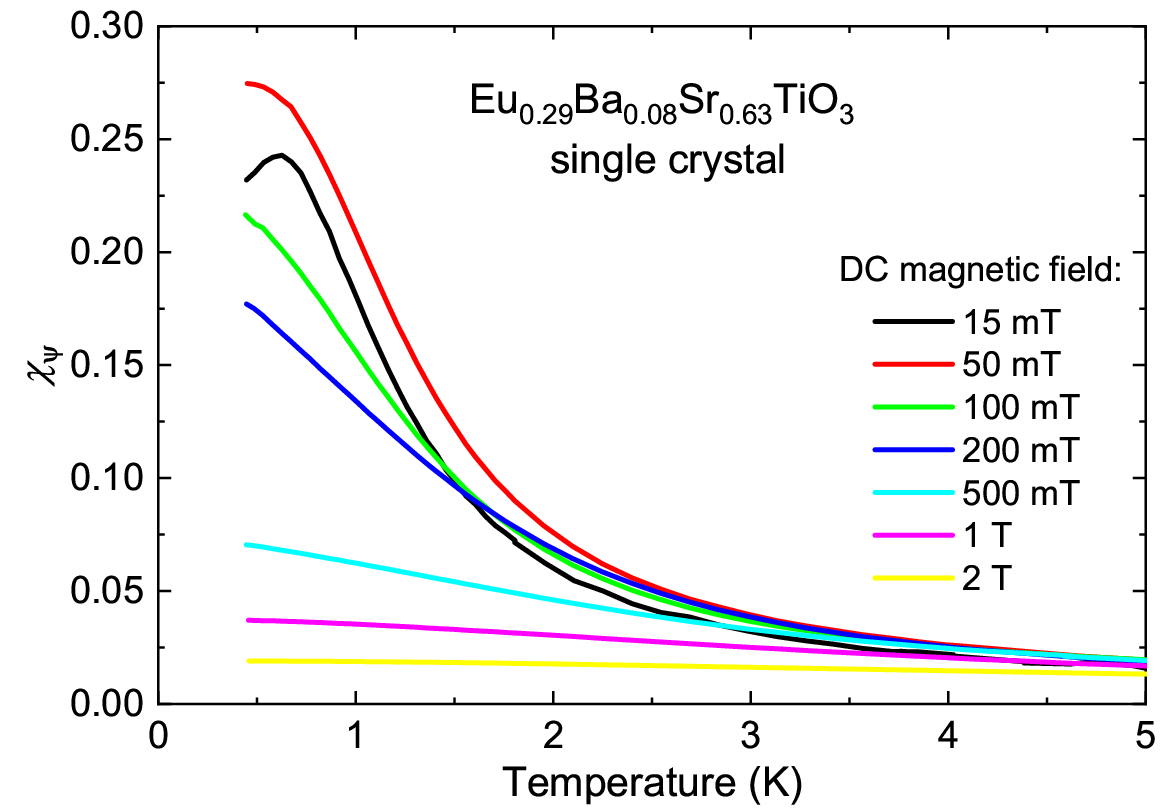}
	\caption{Temperature dependence of magnetic susceptibility measured in various DC magnetic fields on the EBSTO single crystal. The decrease of $\chi_{\boldsymbol{\uppsi}}$ with increasing magnetic field is a consequence of gradual saturation of magnetization (see also Fig. S13 in Supplemental material\cite{Suppl}). At low temperatures, this effect is significant even for weak magnetic fields.}
	\label{fig:magneticka_susceptibilita_single_crystal}
\end{figure}

\begin{figure}[]
	\centering
	\includegraphics[width=84mm]{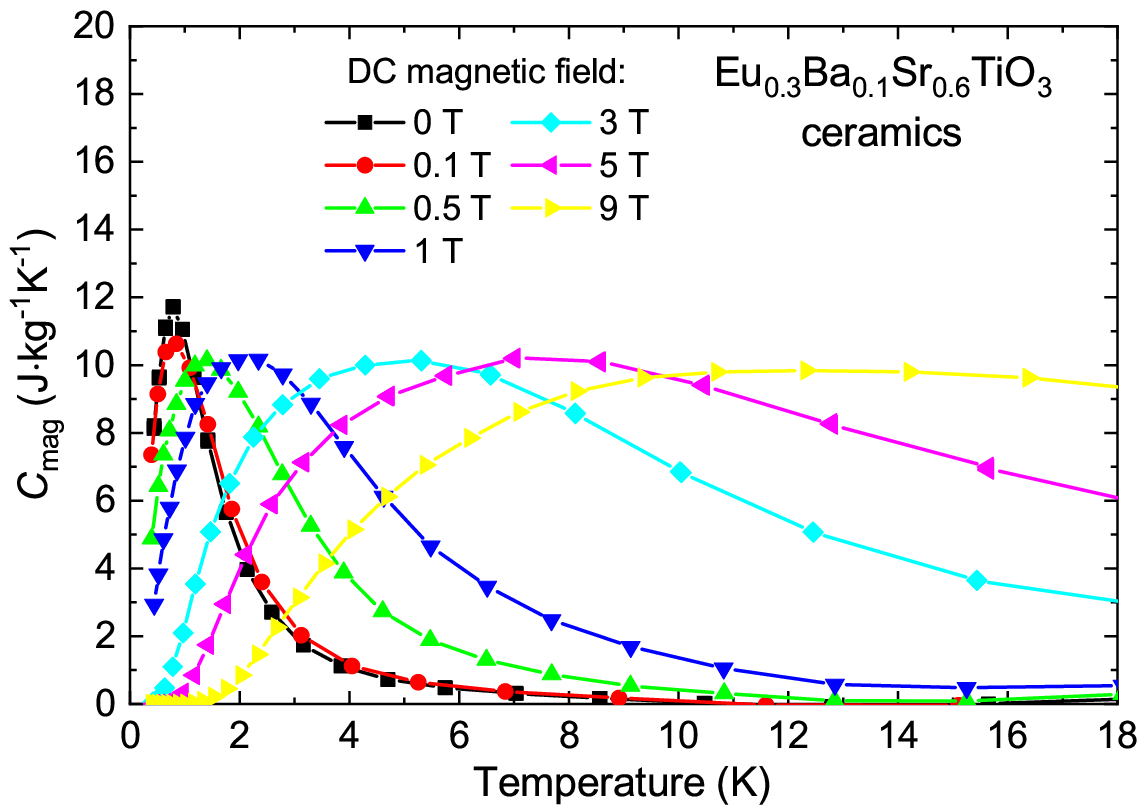}
	\caption{Temperature dependence of the magnetic contribution to heat capacity measured in various DC magnetic fields on the Eu$_{0.3}$Ba$_{0.1}$Sr$_{0.6}$TiO$_3$ ceramics.}
	\label{fig:tepelna_kapacita_stara_keramika}
\end{figure}

Application of the external DC magnetic field causes broadening and shift of the peak in heat capacity towards higher temperatures. We ascribe it to a change of magnetic entropy (saturated paramagnetism), when the mutually aligned magnetic dipoles of Eu$^{2+}$ ions are thermally disordered. Stronger magnetic field implies higher thermal energy required for the disruption of the induced magnetic ordering. Qualitatively similar effect has been observed in pure EuTiO$_3$.\cite{Petrovic13,Jaoui23} The data of $C_{\mathrm{mag}}/T$ published in Ref. \cite{Jaoui23} follow the same magnetic-field dependence as our data presented in Fig. S14. The data of $C_{\mathrm{mag}}$ in Ref. \cite{Petrovic13} are consistent with our Fig. \ref{fig:tepelna_kapacita_stara_keramika} for stronger magnetic field. Nevertheless, the peak in Fig. \ref{fig:tepelna_kapacita_stara_keramika} is not that prominent compared to the Ref. \cite{Petrovic13} in the zero and weak applied magnetic field and also the peak position shifts to higher temperatures with increasing magnetic field monotonously unlike in the Ref. \cite{Petrovic13} and in the data of magnetic susceptibility (Fig. \ref{fig:magneticka_susceptibilita_single_crystal} and Figs. S10, S11 in Supplemental material\cite{Suppl}). We attribute this behavior to the lower concentration of Eu$^{2+}$ ions and very weak spin-flop critical field $<0.1$\,T. However, more detailed investigation of the weak-field $C_{\mathrm{mag}}$ would be necessary for an accurate comparison with the Ref. \cite{Petrovic13} and the AC and DC magnetic susceptibility data. Overall, the zero-field heat capacity peak indeed marks $\Tn$ but the following shift of the peak to higher temperatures is not related to a change of $\Tn$.

Taking into account both the data of heat capacity and the data of DC and AC magnetic susceptibility, we conclude that our EBSTO samples (which should in theory exhibit the AFM QC\cite{Narayan19}) undergo the AFM PT at finite, but very low temperatures below $0.8\K$. Therefore the AFM quantum fluctuations still may play crucial role in the proximity of $\Tn$.

Finally, we turn attention to the temperature dependence of magnetic susceptibility $\chi_{\boldsymbol{\uppsi}} (T)$, which is the main indicator of the possible magnetic QC. It is immediately seen from $\chi_{\boldsymbol{\uppsi}}^{-1} (T)$ measured in DC magnetic field (Fig. \ref{fig:magn_susc_and_inv_magn_susc_on_T} and Fig. S11 in Supplemental material\cite{Suppl}) that both EBSTO single crystal and ceramics exhibit the classical C-W behavior ($\chi_{\boldsymbol{\uppsi}} (T) = \frac{C_{\boldsymbol{\uppsi}}}{T-\theta} + \chi_{\boldsymbol{\uppsi}}^\infty$, where $C_{\boldsymbol{\uppsi}}$ is Curie constant, $\theta$ antiferromagnetic Curie temperature, and $\chi_{\boldsymbol{\uppsi}}^\infty$ possible high-temperature shift) at least down to $4\K$, i.e. very close to $\Tn$. The C-W fits (performed in $4-200\K$ temperature range) give positive antiferromagnetic Curie temperature $\theta = 0.8\K$ for the single crystal and the Eu$_{0.3}$Ba$_{0.1}$Sr$_{0.6}$TiO$_3$ ceramics, while the Eu$_{0.25}$Ba$_{0.1}$Sr$_{0.65}$TiO$_3$ ceramics shows $\theta = 0.3\K$ (which is consistent with positive $\theta = 3.17\K$ of the pure EuTiO$_3$\cite{Katsufuji01} and decreasing concentration of Eu$^{2+}$ ions). These values of $\theta$ are very similar and for single crystal and the Eu$_{0.3}$Ba$_{0.1}$Sr$_{0.6}$TiO$_3$ ceramics possibly even slightly higher than corresponding Néel temperature $\Tn$ of the AFM PT. For the Eu$_{0.25}$Ba$_{0.1}$Sr$_{0.65}$TiO$_3$, $\Tn$ is not known from DC magnetization exactly, since it lies on the verge of our measuring temperature range (Fig. \ref{fig:magn_susc_and_inv_magn_susc_on_T}). The heat capacity data give $\Tn \approx 0.5\K$. However, for the EBSTO single crystal and the Eu$_{0.3}$Ba$_{0.1}$Sr$_{0.6}$TiO$_3$ ceramics, the maximum of $\chi_{\boldsymbol{\uppsi}} (T)$ measured at 15\,mT (Fig. \ref{fig:magn_susc_and_inv_magn_susc_on_T} and Fig. S11 in Supplemental material\cite{Suppl}) is located at $0.6\K$ and $0.7\K$, respectively. Moreover, the maximum of AC magnetic susceptibility (where the applied AC magnetic field is only 0.01\,mT) for the single crystal is at $\Tn = 0.7\K$ (Figs. S10, S12 in Supplemental material\cite{Suppl}), i.e. $0.1\K$ below $\theta = 0.8\K$. This could indicate presence of the AFM QC, since $\chi_{\boldsymbol{\uppsi}} (T)$ does not diverge at $\theta$ and magnetic quantum fluctuations may thus partially suppress the AFM PT. Similarly, SrTiO$_3$ is characterized by a positive Curie temperature of the high-temperature C-W fit,\cite{Weaver59} while it is a critical quantum paraelectric with no FE PT (no divergence of $\chi_{\boldsymbol{\upphi}} (T)$ at this Curie temperature). Nevertheless, the difference $\theta - \Tn \approx 0.1\K$ is marginal and could be only a consequence of measurement error of the temperature sweep.

\begin{figure}[]
	\centering
	\includegraphics[width=83mm]{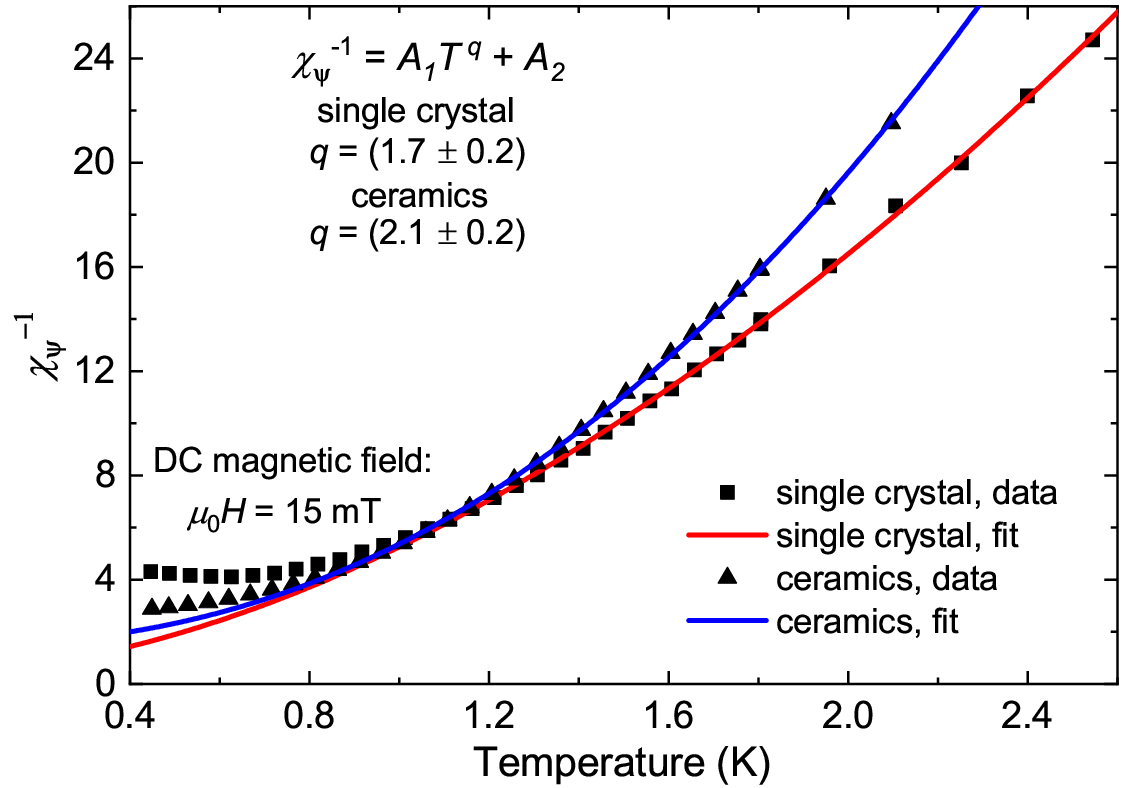}
	\caption{Temperature dependence of inverse DC magnetic susceptibility measured in weak DC magnetic field of 15\,mT on the EBSTO single crystal and Eu$_{0.25}$Ba$_{0.1}$Sr$_{0.65}$TiO$_3$ ceramics. The data of $\chi_{\boldsymbol{\uppsi}}^{-1}$ follow the power-law dependence with parameters $A_1 = (4.8 \pm 0.3)\K^{-q}$, $A_2 = (0.5 \pm 0.3)$, $q = (1.7 \pm 0.2)$ in the temperature range $1.0-2.6\K$ (single crystal) and $A_1 = (3.9 \pm 0.7)\K^{-q}$, $A_2 = (1.5 \pm 0.9)$, $q = (2.1 \pm 0.2)$ in the temperature range $0.8-2.1\K$ (ceramics).}
	\label{fig:inverzni_permeabilita_s_fitem_keramika_monokrystal}
\end{figure}

\begin{figure}[]
	\centering
	\includegraphics[width=87mm]{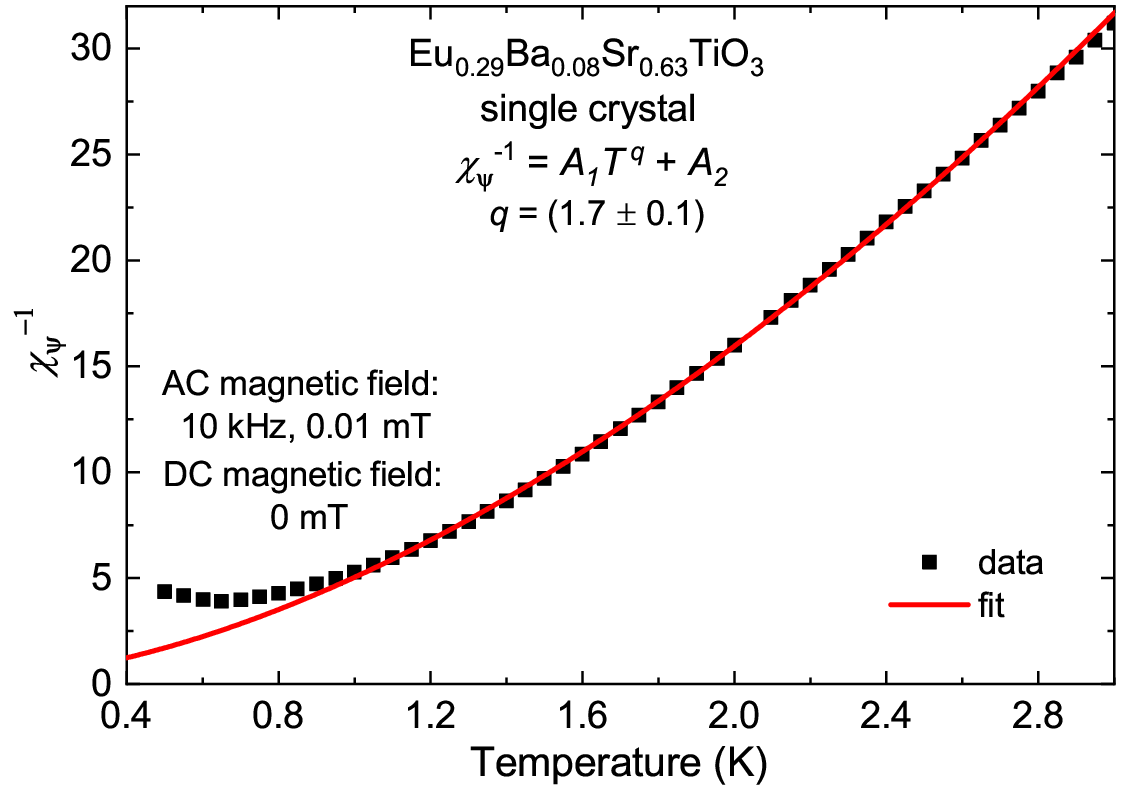}
	\caption{Temperature dependence of inverse AC magnetic susceptibility $\chi_{\boldsymbol{\uppsi}}^{-1}$ measured in zero DC magnetic field on the EBSTO single crystal. The data of $\chi_{\boldsymbol{\uppsi}}^{-1}(T)$ (obtained by normalizing the data of AC $\chi_{\boldsymbol{\uppsi}}(T)$ from Figure S10 to the DC $\chi_{\boldsymbol{\uppsi}}(T)$) follow the power-law dependence with parameters $A_1 = (4.8 \pm 0.2)\K^{-q}$, $A_2 = (0.2 \pm 0.2)$, $q = (1.7 \pm 0.1)$ in the temperature range $1-3\K$.}
	\label{fig:inverzni_AC_permeabilita_s_fitem_monokrystal}
\end{figure}

However, another trace of the AFM QC appears. Closer examination of the DC $\chi_{\boldsymbol{\uppsi}}^{-1} (T)$ data below $3\K$ (Fig. \ref{fig:inverzni_permeabilita_s_fitem_keramika_monokrystal}) reveals that $\chi_{\boldsymbol{\uppsi}}^{-1} (T)$ indeed follows the power-law dependence (see Fig. \ref{fig:inverzni_permeabilita_s_fitem_keramika_monokrystal}) with exponent $q = (1.7 \pm 0.2)$ for the single crystal and $q = (2.1 \pm 0.2)$ for the Eu$_{0.25}$Ba$_{0.1}$Sr$_{0.65}$TiO$_3$ ceramics. The obtained parameters are reliable in the temperature range $1.0-2.6\K$ for the single crystal and $0.8-2.1\K$ for the ceramics, respectively. The rather significant standard deviations already include inaccuracies emerging from variations of the low-temperature limit and length of the fitted temperature range.

The DC $\chi_{\boldsymbol{\uppsi}}^{-1} (T)$ measured in DC magnetic field on the EBSTO single crystal containing 29\,\% of Eu follows power-law dependence with $q = (1.7 \pm 0.2)$. This exponent is similar or somewhat higher than the exponent $3/2$ expected for the AFM QC (and also exponent $4/3$ expected for the eventual weak FM QC) alone.\cite{Narayan19} This could be a trace of the predicted interaction between the FE and AFM quantum fluctuations in the MFQC system. The data of AC $\chi_{\boldsymbol{\uppsi}}^{-1} (T)$ (Fig. \ref{fig:inverzni_AC_permeabilita_s_fitem_monokrystal}) are also promising, as its power-law dependence posses $q = (1.7 \pm 0.1)$ - the same exponent as DC $\chi_{\boldsymbol{\uppsi}}^{-1} (T)$, but determined more accurately and observable in the even slightly broader temperature interval $1-3\K$. Moreover, the DC $\chi_{\boldsymbol{\uppsi}}^{-1} (T)$ of the ceramic sample with 25\,\% of Eu, which should be tuned closer the AFM QC, exhibit power-law dependence with the exponent $q = (2.1 \pm 0.2)$ - value clearly deviating from the standard AFM QC defined by $q = 3/2$ and thus supporting the idea of MFQC behavior. On the other hand, no clear crossover in dielectric susceptibility from $\chi_{\boldsymbol{\upphi}}^{-1}\propto T^{2}$ to different powers was observed, since the temperature range of the possible MFQC (up to $3\K$) coincides with the upturn in $\chi_{\boldsymbol{\upphi}}^{-1}$ caused predominantly by the spin-phonon coupling. Therefore, the type, strength and even actual presence of the interaction is not conclusively confirmed as the FE QC cannot be reliably analyzed below $\approx$ 20\,K.

At a first glance it seems strange that the AFM QC region appears to be so narrow (only up to $3\K$) compared to the FE QC region (up to $60\K$), nonetheless the magnetic quantum fluctuations are often based on magnetic dipolar interaction, which may be similar or dominant over the exchange interaction in quantum critical magnets.\cite{Beauvillain78,Bitko96} Such weak interaction (few orders of magnitude weaker than the electric dipolar one\cite{Chandra17}) then might lead to the quantum critical behavior only at very low temperatures and close proximity of the QCP.\cite{Beauvillain78,Griffin80} Moreover, quantum fluctuations may be significantly reduced, because of the imperfect tuning of the AFM QCP (finite $\Tn$). Several other effects (combined together) could also partially suppress the AFM QC behavior: 1)~rather large magnetic moments of Eu$^{2+}$ ions behave in a more classical way than expected in calculations,\cite{Narayan19} 2) non-homogeneous distribution of Eu$^{2+}$ ions in samples (below the spatial resolution of WDS) weakens their effective interaction, 3) crystal field splitting of Eu$^{2+}$ electronic levels changes character of the exchange interaction, and 4) FE quantum fluctuations interact with the spins in unexpected way and effectively suppress magnetic quantum fluctuations.

\section{Conclusion}

An extensive investigation of magnetic and dielectric properties of the expected multiferroic quantum critical system (Eu,Ba,Sr)TiO$_3$ revealed several noteworthy results. 

The temperature dependence of low-frequency dielectric susceptibility showed ferroelectric quantum criticality in broad temperature range, similarly to pure SrTiO$_3$.\cite{Rowley14} The ferroelectric quantum critical behavior was clearly observed in the temperature range $\approx 20-60\K$ in the single crystal Eu$_{0.29}$Ba$_{0.08}$Sr$_{0.63}$TiO$_3$ sample, which (unlike the ceramic samples) does not contain any significant amount of structural defects. The presence of defects cause additional dielectric relaxations to appear, thus making the temperature-scaling analysis difficult. Upon cooling below $\approx 20\K$, the ferroelectric quantum criticality is initially influenced and then below $10\K$ completely hidden by the predominating spin-phonon coupling.

Measurement of $\chi_{\boldsymbol{\upphi}}$ at various hydrostatic pressures revealed suppression of the ferroelectric quantum critical behavior on pressure increase. This result is consistent with those earlier reported for SrTiO$_3$.

The heat capacity data together with the temperature dependence of both DC and AC magnetic susceptibility ($\chi_{\boldsymbol{\uppsi}} (T)$) confirmed the presence of the AFM PT at finite temperatures from $0.4\K$ to $0.8\K$ for all studied (Eu,Ba,Sr)TiO$_3$ samples. The antiferromagnetic critical point in Eu$_{0.3}$Ba$_{0.1}$Sr$_{0.6}$TiO$_3$ is thus not perfectly tuned to quantum regime ($0\K$) as was calculated earlier.\cite{Narayan19} However, the antiferromagnetic quantum fluctuations are arguably strong enough to produce quantum critical response below $3\K$, where temperature scaling of inverse magnetic susceptibility is changed from classical Curie-Weiss trend to power-law dependence with exponents $q = (1.7 \pm 0.1)$ (AC susceptibility, single crystal, temperature range $1-3\K$), $q = (1.7 \pm 0.2)$ (DC susceptibility, single crystal, temperature range $1.0-2.6\K$), and $q = (2.1 \pm 0.2)$ (DC susceptibility, Eu$_{0.25}$Ba$_{0.1}$Sr$_{0.65}$TiO$_3$ ceramics, temperature range $0.8-2.1\K$), i.e. values deviating from the antiferromagnetic quantum critical exponent $q = 3/2$. Therefore we the multiferroic quantum critical behavior to be present in (Eu,Ba,Sr)TiO$_3$ system (assuming that the ferroelectric quantum criticality is not suppressed by the spin-phonon coupling below $10\K$) as expected from calculations. Nevertheless, in comparison with the ferroelectric quantum critical behavior ($\chi_{\boldsymbol{\upphi}}^{-1}\propto T^{2}$), which is present up to $60\K$, the antiferromagnetic quantum fluctuations affects the $\chi_{\boldsymbol{\uppsi}}^{-1} (T)$ scaling only below $3\K$ - only slightly above the Néel temperature located (in the single crystal) at $\Tn = 0.7\K$. This is probably due to both weakness of the exchange and magnetic dipolar interaction compared to the electric dipolar one and imperfect tuning of the antiferromagnetic quantum critical point, even though several other effects may also be at play.

Comparison of exponents in power-law dependences of the inverse magnetic susceptibility $q = (1.7 \pm 0.1)$ ($q = (1.7 \pm 0.2)$) and $q = (2.1 \pm 0.2)$ for the single crystal and the Eu$_{0.25}$Ba$_{0.1}$Sr$_{0.65}$TiO$_3$ ceramics, respectively, gives a trace of stronger interaction between ferroelectric and antiferromagnetic quantum fluctuations in the Eu$_{0.25}$Ba$_{0.1}$Sr$_{0.65}$TiO$_3$ ceramic sample, which is tuned closer to the antiferromagnetic quantum criticality. However, we did not observe any serious indications of crossover from $\chi_{\boldsymbol{\upphi}}^{-1}\propto T^{2}$ to different temperature scalings of $\chi_{\boldsymbol{\upphi}}^{-1}$ (as theory predicts for mutually interacting quantum fluctuations in multiferroic quantum critical systems\cite{Narayan19}) below $3\K$, because of the spin-phonon coupling, which predominates below $10\K$. Therefore, it is not possible to analyze the strength and type of the prevailing interaction between the ferroelectric and antiferromagnetic quantum fluctuations in (Eu,Ba,Sr)TiO$_3$ solid solution and more work on other (Eu,Ba,Sr)TiO$_3$ compositions is required to confirm the MFQC unambiguously.

\begin{acknowledgments}
This work has been supported by the Czech Science Foundation (Project No. 21-06802S) and the Grant Agency of the Czech Technical University in Prague (Project No. SGS22/182/OHK4/3T/14). Experiments were also performed in MGML (mgml.eu), which is supported within the program of Czech Research Infrastructures (project no. LM2023065).

We would like to acknowledge Dr. Karel Jurek and Dr. Mark\'{e}ta Jaro\v{s}ov\'{a} for careful WDS analysis of our samples, Dr. Pavel Nov\'{a}k for calculation of possible crystal field effects, and Prof. Vladim\'{i}r Sechovsk\'{y} for helpful discussion on interpretation of our magnetic data.

\end{acknowledgments}


\begin{thebibliography}{53}%
	\makeatletter
	\providecommand \@ifxundefined [1]{%
		\@ifx{#1\undefined}
	}%
	\providecommand \@ifnum [1]{%
		\ifnum #1\expandafter \@firstoftwo
		\else \expandafter \@secondoftwo
		\fi
	}%
	\providecommand \@ifx [1]{%
		\ifx #1\expandafter \@firstoftwo
		\else \expandafter \@secondoftwo
		\fi
	}%
	\providecommand \natexlab [1]{#1}%
	\providecommand \enquote  [1]{``#1''}%
	\providecommand \bibnamefont  [1]{#1}%
	\providecommand \bibfnamefont [1]{#1}%
	\providecommand \citenamefont [1]{#1}%
	\providecommand \href@noop [0]{\@secondoftwo}%
	\providecommand \href [0]{\begingroup \@sanitize@url \@href}%
	\providecommand \@href[1]{\@@startlink{#1}\@@href}%
	\providecommand \@@href[1]{\endgroup#1\@@endlink}%
	\providecommand \@sanitize@url [0]{\catcode `\\12\catcode `\$12\catcode
		`\&12\catcode `\#12\catcode `\^12\catcode `\_12\catcode `\%12\relax}%
	\providecommand \@@startlink[1]{}%
	\providecommand \@@endlink[0]{}%
	\providecommand \url  [0]{\begingroup\@sanitize@url \@url }%
	\providecommand \@url [1]{\endgroup\@href {#1}{\urlprefix }}%
	\providecommand \urlprefix  [0]{URL }%
	\providecommand \Eprint [0]{\href }%
	\providecommand \doibase [0]{https://doi.org/}%
	\providecommand \selectlanguage [0]{\@gobble}%
	\providecommand \bibinfo  [0]{\@secondoftwo}%
	\providecommand \bibfield  [0]{\@secondoftwo}%
	\providecommand \translation [1]{[#1]}%
	\providecommand \BibitemOpen [0]{}%
	\providecommand \bibitemStop [0]{}%
	\providecommand \bibitemNoStop [0]{.\EOS\space}%
	\providecommand \EOS [0]{\spacefactor3000\relax}%
	\providecommand \BibitemShut  [1]{\csname bibitem#1\endcsname}%
	\let\auto@bib@innerbib\@empty
	\bibitem [{\citenamefont {Sachdev}(2011)}]{Sachdev11}%
	\BibitemOpen
	\bibfield  {author} {\bibinfo {author} {\bibfnamefont {S.}~\bibnamefont
			{Sachdev}},\ }\href {https://doi.org/10.1017/CBO9780511973765} {\emph
		{\bibinfo {title} {Quantum Phase Transitions}}},\ \bibinfo {edition} {2nd}\
	ed.\ (\bibinfo  {publisher} {Cambridge University Press},\ \bibinfo {year}
	{2011})\BibitemShut {NoStop}%
	\bibitem [{\citenamefont {Gegenwart}\ \emph {et~al.}(2008)\citenamefont
		{Gegenwart}, \citenamefont {Si},\ and\ \citenamefont
		{Steglich}}]{Gegenwart08}%
	\BibitemOpen
	\bibfield  {author} {\bibinfo {author} {\bibfnamefont {P.}~\bibnamefont
			{Gegenwart}}, \bibinfo {author} {\bibfnamefont {Q.}~\bibnamefont {Si}},\ and\
		\bibinfo {author} {\bibfnamefont {F.}~\bibnamefont {Steglich}},\ }\bibfield
	{title} {\bibinfo {title} {Quantum criticality in heavy-fermion metals},\
	}\href@noop {} {\bibfield  {journal} {\bibinfo  {journal} {Nature Phys.}\
		}\textbf {\bibinfo {volume} {4}},\ \bibinfo {pages} {186} (\bibinfo {year}
		{2008})}\BibitemShut {NoStop}%
	\bibitem [{\citenamefont {Mathur}\ \emph {et~al.}(1998)\citenamefont {Mathur},
		\citenamefont {Grosche}, \citenamefont {Julian}, \citenamefont {Walker},
		\citenamefont {Freye}, \citenamefont {Haselwimmer},\ and\ \citenamefont
		{Lonzarich}}]{Mathur98}%
	\BibitemOpen
	\bibfield  {author} {\bibinfo {author} {\bibfnamefont {N.}~\bibnamefont
			{Mathur}}, \bibinfo {author} {\bibfnamefont {F.}~\bibnamefont {Grosche}},
		\bibinfo {author} {\bibfnamefont {S.}~\bibnamefont {Julian}}, \bibinfo
		{author} {\bibfnamefont {I.}~\bibnamefont {Walker}}, \bibinfo {author}
		{\bibfnamefont {D.}~\bibnamefont {Freye}}, \bibinfo {author} {\bibfnamefont
			{R.}~\bibnamefont {Haselwimmer}},\ and\ \bibinfo {author} {\bibfnamefont
			{G.}~\bibnamefont {Lonzarich}},\ }\bibfield  {title} {\bibinfo {title}
		{Magnetically mediated superconductivity in heavy fermion compounds},\
	}\href@noop {} {\bibfield  {journal} {\bibinfo  {journal} {Nature}\ }\textbf
		{\bibinfo {volume} {394}},\ \bibinfo {pages} {39} (\bibinfo {year}
		{1998})}\BibitemShut {NoStop}%
	\bibitem [{\citenamefont {Saxena}\ \emph {et~al.}(2000)\citenamefont {Saxena},
		\citenamefont {Agarwal}, \citenamefont {Ahilan}, \citenamefont {Grosche},
		\citenamefont {Haselwimmer}, \citenamefont {Steiner}, \citenamefont {Pugh},
		\citenamefont {Walker}, \citenamefont {Julian}, \citenamefont {Monthoux}
		\emph {et~al.}}]{Saxena00}%
	\BibitemOpen
	\bibfield  {author} {\bibinfo {author} {\bibfnamefont {S.}~\bibnamefont
			{Saxena}}, \bibinfo {author} {\bibfnamefont {P.}~\bibnamefont {Agarwal}},
		\bibinfo {author} {\bibfnamefont {K.}~\bibnamefont {Ahilan}}, \bibinfo
		{author} {\bibfnamefont {F.}~\bibnamefont {Grosche}}, \bibinfo {author}
		{\bibfnamefont {R.}~\bibnamefont {Haselwimmer}}, \bibinfo {author}
		{\bibfnamefont {M.}~\bibnamefont {Steiner}}, \bibinfo {author} {\bibfnamefont
			{E.}~\bibnamefont {Pugh}}, \bibinfo {author} {\bibfnamefont {I.}~\bibnamefont
			{Walker}}, \bibinfo {author} {\bibfnamefont {S.}~\bibnamefont {Julian}},
		\bibinfo {author} {\bibfnamefont {P.}~\bibnamefont {Monthoux}}, \emph
		{et~al.},\ }\bibfield  {title} {\bibinfo {title} {Superconductivity on the
			border of itinerant-electron ferromagnetism in {UGe$_2$}},\ }\href@noop {}
	{\bibfield  {journal} {\bibinfo  {journal} {Nature}\ }\textbf {\bibinfo
			{volume} {406}},\ \bibinfo {pages} {587} (\bibinfo {year}
		{2000})}\BibitemShut {NoStop}%
	\bibitem [{\citenamefont {Ran}\ \emph {et~al.}(2019)\citenamefont {Ran},
		\citenamefont {Eckberg}, \citenamefont {Ding}, \citenamefont {Furukawa},
		\citenamefont {Metz}, \citenamefont {Saha}, \citenamefont {Liu},
		\citenamefont {Zic}, \citenamefont {Kim}, \citenamefont {Paglione} \emph
		{et~al.}}]{Ran19}%
	\BibitemOpen
	\bibfield  {author} {\bibinfo {author} {\bibfnamefont {S.}~\bibnamefont
			{Ran}}, \bibinfo {author} {\bibfnamefont {C.}~\bibnamefont {Eckberg}},
		\bibinfo {author} {\bibfnamefont {Q.-P.}\ \bibnamefont {Ding}}, \bibinfo
		{author} {\bibfnamefont {Y.}~\bibnamefont {Furukawa}}, \bibinfo {author}
		{\bibfnamefont {T.}~\bibnamefont {Metz}}, \bibinfo {author} {\bibfnamefont
			{S.~R.}\ \bibnamefont {Saha}}, \bibinfo {author} {\bibfnamefont {I.-L.}\
			\bibnamefont {Liu}}, \bibinfo {author} {\bibfnamefont {M.}~\bibnamefont
			{Zic}}, \bibinfo {author} {\bibfnamefont {H.}~\bibnamefont {Kim}}, \bibinfo
		{author} {\bibfnamefont {J.}~\bibnamefont {Paglione}}, \emph {et~al.},\
	}\bibfield  {title} {\bibinfo {title} {Nearly ferromagnetic spin-triplet
			superconductivity},\ }\href@noop {} {\bibfield  {journal} {\bibinfo
			{journal} {Science}\ }\textbf {\bibinfo {volume} {365}},\ \bibinfo {pages}
		{684} (\bibinfo {year} {2019})}\BibitemShut {NoStop}%
	\bibitem [{\citenamefont {Lake}\ \emph {et~al.}(2005)\citenamefont {Lake},
		\citenamefont {Tennant}, \citenamefont {Frost},\ and\ \citenamefont
		{Nagler}}]{Lake05}%
	\BibitemOpen
	\bibfield  {author} {\bibinfo {author} {\bibfnamefont {B.}~\bibnamefont
			{Lake}}, \bibinfo {author} {\bibfnamefont {D.~A.}\ \bibnamefont {Tennant}},
		\bibinfo {author} {\bibfnamefont {C.~D.}\ \bibnamefont {Frost}},\ and\
		\bibinfo {author} {\bibfnamefont {S.~E.}\ \bibnamefont {Nagler}},\ }\bibfield
	{title} {\bibinfo {title} {Quantum criticality and universal scaling of a
			quantum antiferromagnet},\ }\href@noop {} {\bibfield  {journal} {\bibinfo
			{journal} {Nature Mat.}\ }\textbf {\bibinfo {volume} {4}},\ \bibinfo {pages}
		{329} (\bibinfo {year} {2005})}\BibitemShut {NoStop}%
	\bibitem [{\citenamefont {Rowley}\ \emph {et~al.}(2014)\citenamefont {Rowley},
		\citenamefont {Spalek}, \citenamefont {Smith}, \citenamefont {Dean},
		\citenamefont {Itoh}, \citenamefont {Scott}, \citenamefont {Lonzarich},\ and\
		\citenamefont {Saxena}}]{Rowley14}%
	\BibitemOpen
	\bibfield  {author} {\bibinfo {author} {\bibfnamefont {S.}~\bibnamefont
			{Rowley}}, \bibinfo {author} {\bibfnamefont {L.}~\bibnamefont {Spalek}},
		\bibinfo {author} {\bibfnamefont {R.}~\bibnamefont {Smith}}, \bibinfo
		{author} {\bibfnamefont {M.}~\bibnamefont {Dean}}, \bibinfo {author}
		{\bibfnamefont {M.}~\bibnamefont {Itoh}}, \bibinfo {author} {\bibfnamefont
			{J.}~\bibnamefont {Scott}}, \bibinfo {author} {\bibfnamefont
			{G.}~\bibnamefont {Lonzarich}},\ and\ \bibinfo {author} {\bibfnamefont
			{S.}~\bibnamefont {Saxena}},\ }\bibfield  {title} {\bibinfo {title}
		{Ferroelectric quantum criticality},\ }\href@noop {} {\bibfield  {journal}
		{\bibinfo  {journal} {Nature Phys.}\ }\textbf {\bibinfo {volume} {10}},\
		\bibinfo {pages} {367} (\bibinfo {year} {2014})}\BibitemShut {NoStop}%
	\bibitem [{\citenamefont {Roussev}\ and\ \citenamefont
		{Millis}(2003)}]{Roussev03}%
	\BibitemOpen
	\bibfield  {author} {\bibinfo {author} {\bibfnamefont {R.}~\bibnamefont
			{Roussev}}\ and\ \bibinfo {author} {\bibfnamefont {A.}~\bibnamefont
			{Millis}},\ }\bibfield  {title} {\bibinfo {title} {Theory of the quantum
			paraelectric-ferroelectric transition},\ }\href@noop {} {\bibfield  {journal}
		{\bibinfo  {journal} {Phys. Rev. B}\ }\textbf {\bibinfo {volume} {67}},\
		\bibinfo {pages} {014105} (\bibinfo {year} {2003})}\BibitemShut {NoStop}%
	\bibitem [{\citenamefont {Edge}\ \emph {et~al.}(2015)\citenamefont {Edge},
		\citenamefont {Kedem}, \citenamefont {Aschauer}, \citenamefont {Spaldin},\
		and\ \citenamefont {Balatsky}}]{Edge15}%
	\BibitemOpen
	\bibfield  {author} {\bibinfo {author} {\bibfnamefont {J.~M.}\ \bibnamefont
			{Edge}}, \bibinfo {author} {\bibfnamefont {Y.}~\bibnamefont {Kedem}},
		\bibinfo {author} {\bibfnamefont {U.}~\bibnamefont {Aschauer}}, \bibinfo
		{author} {\bibfnamefont {N.~A.}\ \bibnamefont {Spaldin}},\ and\ \bibinfo
		{author} {\bibfnamefont {A.~V.}\ \bibnamefont {Balatsky}},\ }\bibfield
	{title} {\bibinfo {title} {Quantum critical origin of the superconducting
			dome in {SrTiO$_3$}},\ }\href@noop {} {\bibfield  {journal} {\bibinfo
			{journal} {Phys. Rev. Lett.}\ }\textbf {\bibinfo {volume} {115}},\ \bibinfo
		{pages} {247002} (\bibinfo {year} {2015})}\BibitemShut {NoStop}%
	\bibitem [{\citenamefont {Rowley}\ \emph {et~al.}(2010)\citenamefont {Rowley},
		\citenamefont {Smith}, \citenamefont {Dean}, \citenamefont {Spalek},
		\citenamefont {Sutherland}, \citenamefont {Saxena}, \citenamefont {Alireza},
		\citenamefont {Ko}, \citenamefont {Liu}, \citenamefont {Pugh} \emph
		{et~al.}}]{Rowley10}%
	\BibitemOpen
	\bibfield  {author} {\bibinfo {author} {\bibfnamefont {S.}~\bibnamefont
			{Rowley}}, \bibinfo {author} {\bibfnamefont {R.}~\bibnamefont {Smith}},
		\bibinfo {author} {\bibfnamefont {M.}~\bibnamefont {Dean}}, \bibinfo {author}
		{\bibfnamefont {L.}~\bibnamefont {Spalek}}, \bibinfo {author} {\bibfnamefont
			{M.}~\bibnamefont {Sutherland}}, \bibinfo {author} {\bibfnamefont
			{M.}~\bibnamefont {Saxena}}, \bibinfo {author} {\bibfnamefont
			{P.}~\bibnamefont {Alireza}}, \bibinfo {author} {\bibfnamefont
			{C.}~\bibnamefont {Ko}}, \bibinfo {author} {\bibfnamefont {C.}~\bibnamefont
			{Liu}}, \bibinfo {author} {\bibfnamefont {E.}~\bibnamefont {Pugh}}, \emph
		{et~al.},\ }\bibfield  {title} {\bibinfo {title} {Ferromagnetic and
			ferroelectric quantum phase transitions},\ }\href@noop {} {\bibfield
		{journal} {\bibinfo  {journal} {Phys. Status Solidi B}\ }\textbf {\bibinfo
			{volume} {247}},\ \bibinfo {pages} {469} (\bibinfo {year}
		{2010})}\BibitemShut {NoStop}%
	\bibitem [{\citenamefont {Narayan}\ \emph {et~al.}(2019)\citenamefont
		{Narayan}, \citenamefont {Cano}, \citenamefont {Balatsky},\ and\
		\citenamefont {Spaldin}}]{Narayan19}%
	\BibitemOpen
	\bibfield  {author} {\bibinfo {author} {\bibfnamefont {A.}~\bibnamefont
			{Narayan}}, \bibinfo {author} {\bibfnamefont {A.}~\bibnamefont {Cano}},
		\bibinfo {author} {\bibfnamefont {A.~V.}\ \bibnamefont {Balatsky}},\ and\
		\bibinfo {author} {\bibfnamefont {N.~A.}\ \bibnamefont {Spaldin}},\
	}\bibfield  {title} {\bibinfo {title} {Multiferroic quantum criticality},\
	}\href@noop {} {\bibfield  {journal} {\bibinfo  {journal} {Nature Mat.}\
		}\textbf {\bibinfo {volume} {18}},\ \bibinfo {pages} {223} (\bibinfo {year}
		{2019})}\BibitemShut {NoStop}%
	\bibitem [{\citenamefont {She}\ \emph {et~al.}(2010)\citenamefont {She},
		\citenamefont {Zaanen}, \citenamefont {Bishop},\ and\ \citenamefont
		{Balatsky}}]{She10}%
	\BibitemOpen
	\bibfield  {author} {\bibinfo {author} {\bibfnamefont {J.-H.}\ \bibnamefont
			{She}}, \bibinfo {author} {\bibfnamefont {J.}~\bibnamefont {Zaanen}},
		\bibinfo {author} {\bibfnamefont {A.~R.}\ \bibnamefont {Bishop}},\ and\
		\bibinfo {author} {\bibfnamefont {A.~V.}\ \bibnamefont {Balatsky}},\
	}\bibfield  {title} {\bibinfo {title} {Stability of quantum critical points
			in the presence of competing orders},\ }\href@noop {} {\bibfield  {journal}
		{\bibinfo  {journal} {Phys. Rev. B}\ }\textbf {\bibinfo {volume} {82}},\
		\bibinfo {pages} {165128} (\bibinfo {year} {2010})}\BibitemShut {NoStop}%
	\bibitem [{\citenamefont {Morice}\ \emph {et~al.}(2017)\citenamefont {Morice},
		\citenamefont {Chandra}, \citenamefont {Rowley}, \citenamefont {Lonzarich},\
		and\ \citenamefont {Saxena}}]{Morice17}%
	\BibitemOpen
	\bibfield  {author} {\bibinfo {author} {\bibfnamefont {C.}~\bibnamefont
			{Morice}}, \bibinfo {author} {\bibfnamefont {P.}~\bibnamefont {Chandra}},
		\bibinfo {author} {\bibfnamefont {S.~E.}\ \bibnamefont {Rowley}}, \bibinfo
		{author} {\bibfnamefont {G.}~\bibnamefont {Lonzarich}},\ and\ \bibinfo
		{author} {\bibfnamefont {S.~S.}\ \bibnamefont {Saxena}},\ }\bibfield  {title}
	{\bibinfo {title} {Hidden fluctuations close to a quantum bicritical point},\
	}\href@noop {} {\bibfield  {journal} {\bibinfo  {journal} {Phys. Rev. B}\
		}\textbf {\bibinfo {volume} {96}},\ \bibinfo {pages} {245104} (\bibinfo
		{year} {2017})}\BibitemShut {NoStop}%
	\bibitem [{\citenamefont {Chandra}\ \emph {et~al.}(2017)\citenamefont
		{Chandra}, \citenamefont {Lonzarich}, \citenamefont {Rowley},\ and\
		\citenamefont {Scott}}]{Chandra17}%
	\BibitemOpen
	\bibfield  {author} {\bibinfo {author} {\bibfnamefont {P.}~\bibnamefont
			{Chandra}}, \bibinfo {author} {\bibfnamefont {G.~G.}\ \bibnamefont
			{Lonzarich}}, \bibinfo {author} {\bibfnamefont {S.}~\bibnamefont {Rowley}},\
		and\ \bibinfo {author} {\bibfnamefont {J.}~\bibnamefont {Scott}},\ }\bibfield
	{title} {\bibinfo {title} {Prospects and applications near ferroelectric
			quantum phase transitions: a key issues review},\ }\href@noop {} {\bibfield
		{journal} {\bibinfo  {journal} {Rep. Prog. Phys.}\ }\textbf {\bibinfo
			{volume} {80}},\ \bibinfo {pages} {112502} (\bibinfo {year}
		{2017})}\BibitemShut {NoStop}%
	\bibitem [{\citenamefont {Flavi{\'a}n}\ \emph {et~al.}(2023)\citenamefont
		{Flavi{\'a}n}, \citenamefont {Volkov}, \citenamefont {Hayashida},
		\citenamefont {Povarov}, \citenamefont {Gvasaliya}, \citenamefont {Chandra},\
		and\ \citenamefont {Zheludev}}]{Flavian23}%
	\BibitemOpen
	\bibfield  {author} {\bibinfo {author} {\bibfnamefont {D.}~\bibnamefont
			{Flavi{\'a}n}}, \bibinfo {author} {\bibfnamefont {P.~A.}\ \bibnamefont
			{Volkov}}, \bibinfo {author} {\bibfnamefont {S.}~\bibnamefont {Hayashida}},
		\bibinfo {author} {\bibfnamefont {K.~Y.}\ \bibnamefont {Povarov}}, \bibinfo
		{author} {\bibfnamefont {S.}~\bibnamefont {Gvasaliya}}, \bibinfo {author}
		{\bibfnamefont {P.}~\bibnamefont {Chandra}},\ and\ \bibinfo {author}
		{\bibfnamefont {A.}~\bibnamefont {Zheludev}},\ }\bibfield  {title} {\bibinfo
		{title} {Dielectric relaxation by quantum critical magnons},\ }\href@noop {}
	{\bibfield  {journal} {\bibinfo  {journal} {Phys. Rev. Lett.}\ }\textbf
		{\bibinfo {volume} {130}},\ \bibinfo {pages} {216501} (\bibinfo {year}
		{2023})}\BibitemShut {NoStop}%
	\bibitem [{\citenamefont {Brous}\ \emph {et~al.}(1953)\citenamefont {Brous},
		\citenamefont {Fankuchen},\ and\ \citenamefont {Banks}}]{Brous53}%
	\BibitemOpen
	\bibfield  {author} {\bibinfo {author} {\bibfnamefont {J.}~\bibnamefont
			{Brous}}, \bibinfo {author} {\bibfnamefont {I.}~\bibnamefont {Fankuchen}},\
		and\ \bibinfo {author} {\bibfnamefont {E.}~\bibnamefont {Banks}},\ }\bibfield
	{title} {\bibinfo {title} {Rare earth titanates with a perovskite
			structure},\ }\href@noop {} {\bibfield  {journal} {\bibinfo  {journal} {Acta
				Cryst.}\ }\textbf {\bibinfo {volume} {6}},\ \bibinfo {pages} {67} (\bibinfo
		{year} {1953})}\BibitemShut {NoStop}%
	\bibitem [{\citenamefont {M{\"u}ller}\ \emph {et~al.}(1968)\citenamefont
		{M{\"u}ller}, \citenamefont {Berlinger},\ and\ \citenamefont
		{Waldner}}]{Muller68}%
	\BibitemOpen
	\bibfield  {author} {\bibinfo {author} {\bibfnamefont {K.}~\bibnamefont
			{M{\"u}ller}}, \bibinfo {author} {\bibfnamefont {W.}~\bibnamefont
			{Berlinger}},\ and\ \bibinfo {author} {\bibfnamefont {F.}~\bibnamefont
			{Waldner}},\ }\bibfield  {title} {\bibinfo {title} {Characteristic structural
			phase transition in perovskite-type compounds},\ }\href@noop {} {\bibfield
		{journal} {\bibinfo  {journal} {Phys. Rev. Lett.}\ }\textbf {\bibinfo
			{volume} {21}},\ \bibinfo {pages} {814} (\bibinfo {year} {1968})}\BibitemShut
	{NoStop}%
	\bibitem [{\citenamefont {Fleury}\ \emph {et~al.}(1968)\citenamefont {Fleury},
		\citenamefont {Scott},\ and\ \citenamefont {Worlock}}]{Fleury68}%
	\BibitemOpen
	\bibfield  {author} {\bibinfo {author} {\bibfnamefont {P.}~\bibnamefont
			{Fleury}}, \bibinfo {author} {\bibfnamefont {J.}~\bibnamefont {Scott}},\ and\
		\bibinfo {author} {\bibfnamefont {J.}~\bibnamefont {Worlock}},\ }\bibfield
	{title} {\bibinfo {title} {Soft phonon modes and the 110 {K} phase transition
			in {SrTiO$_3$}},\ }\href@noop {} {\bibfield  {journal} {\bibinfo  {journal}
			{Phys. Rev. Lett.}\ }\textbf {\bibinfo {volume} {21}},\ \bibinfo {pages} {16}
		(\bibinfo {year} {1968})}\BibitemShut {NoStop}%
	\bibitem [{\citenamefont {Weaver}(1959)}]{Weaver59}%
	\BibitemOpen
	\bibfield  {author} {\bibinfo {author} {\bibfnamefont {H.}~\bibnamefont
			{Weaver}},\ }\bibfield  {title} {\bibinfo {title} {Dielectric properties of
			single crystals of {SrTiO$_3$} at low temperatures},\ }\href@noop {}
	{\bibfield  {journal} {\bibinfo  {journal} {J. Phys. Chem. Solids}\ }\textbf
		{\bibinfo {volume} {11}},\ \bibinfo {pages} {274} (\bibinfo {year}
		{1959})}\BibitemShut {NoStop}%
	\bibitem [{\citenamefont {M{\"u}ller}\ and\ \citenamefont
		{Burkard}(1979)}]{Muller79}%
	\BibitemOpen
	\bibfield  {author} {\bibinfo {author} {\bibfnamefont {K.~A.}\ \bibnamefont
			{M{\"u}ller}}\ and\ \bibinfo {author} {\bibfnamefont {H.}~\bibnamefont
			{Burkard}},\ }\bibfield  {title} {\bibinfo {title} {{SrTiO$_3$}: An intrinsic
			quantum paraelectric below 4 {K}},\ }\href@noop {} {\bibfield  {journal}
		{\bibinfo  {journal} {Phys. Rev. B}\ }\textbf {\bibinfo {volume} {19}},\
		\bibinfo {pages} {3593} (\bibinfo {year} {1979})}\BibitemShut {NoStop}%
	\bibitem [{\citenamefont {Bussmann-Holder}\ \emph {et~al.}(2011)\citenamefont
		{Bussmann-Holder}, \citenamefont {K{\"o}hler}, \citenamefont {Kremer},\ and\
		\citenamefont {Law}}]{Bussmann-Holder11}%
	\BibitemOpen
	\bibfield  {author} {\bibinfo {author} {\bibfnamefont {A.}~\bibnamefont
			{Bussmann-Holder}}, \bibinfo {author} {\bibfnamefont {J.}~\bibnamefont
			{K{\"o}hler}}, \bibinfo {author} {\bibfnamefont {R.}~\bibnamefont {Kremer}},\
		and\ \bibinfo {author} {\bibfnamefont {J.}~\bibnamefont {Law}},\ }\bibfield
	{title} {\bibinfo {title} {Relation between structural instabilities in
			{EuTiO$_3$} and {SrTiO$_3$}},\ }\href@noop {} {\bibfield  {journal} {\bibinfo
			{journal} {Phys. Rev. B}\ }\textbf {\bibinfo {volume} {83}},\ \bibinfo
		{pages} {212102} (\bibinfo {year} {2011})}\BibitemShut {NoStop}%
	\bibitem [{\citenamefont {Goian}\ \emph {et~al.}(2012)\citenamefont {Goian},
		\citenamefont {Kamba}, \citenamefont {Pacherov{\'a}}, \citenamefont
		{Drahokoupil}, \citenamefont {Palatinus}, \citenamefont {Du{\v{s}}ek},
		\citenamefont {Rohl{\'\i}{\v{c}}ek}, \citenamefont {Savinov}, \citenamefont
		{Laufek}, \citenamefont {Schranz} \emph {et~al.}}]{Goian12a}%
	\BibitemOpen
	\bibfield  {author} {\bibinfo {author} {\bibfnamefont {V.}~\bibnamefont
			{Goian}}, \bibinfo {author} {\bibfnamefont {S.}~\bibnamefont {Kamba}},
		\bibinfo {author} {\bibfnamefont {O.}~\bibnamefont {Pacherov{\'a}}}, \bibinfo
		{author} {\bibfnamefont {J.}~\bibnamefont {Drahokoupil}}, \bibinfo {author}
		{\bibfnamefont {L.}~\bibnamefont {Palatinus}}, \bibinfo {author}
		{\bibfnamefont {M.}~\bibnamefont {Du{\v{s}}ek}}, \bibinfo {author}
		{\bibfnamefont {J.}~\bibnamefont {Rohl{\'\i}{\v{c}}ek}}, \bibinfo {author}
		{\bibfnamefont {M.}~\bibnamefont {Savinov}}, \bibinfo {author} {\bibfnamefont
			{F.}~\bibnamefont {Laufek}}, \bibinfo {author} {\bibfnamefont
			{W.}~\bibnamefont {Schranz}}, \emph {et~al.},\ }\bibfield  {title} {\bibinfo
		{title} {Antiferrodistortive phase transition in {EuTiO$_3$}},\ }\href@noop
	{} {\bibfield  {journal} {\bibinfo  {journal} {Phys. Rev. B}\ }\textbf
		{\bibinfo {volume} {86}},\ \bibinfo {pages} {054112} (\bibinfo {year}
		{2012})}\BibitemShut {NoStop}%
	\bibitem [{\citenamefont {K{\"o}hler}\ \emph {et~al.}(2012)\citenamefont
		{K{\"o}hler}, \citenamefont {Dinnebier},\ and\ \citenamefont
		{Bussmann-Holder}}]{Kohler12}%
	\BibitemOpen
	\bibfield  {author} {\bibinfo {author} {\bibfnamefont {J.}~\bibnamefont
			{K{\"o}hler}}, \bibinfo {author} {\bibfnamefont {R.}~\bibnamefont
			{Dinnebier}},\ and\ \bibinfo {author} {\bibfnamefont {A.}~\bibnamefont
			{Bussmann-Holder}},\ }\bibfield  {title} {\bibinfo {title} {Structural
			instability of {EuTiO3} from {X}-ray powder diffraction},\ }\href@noop {}
	{\bibfield  {journal} {\bibinfo  {journal} {Phase Trans.}\ }\textbf {\bibinfo
			{volume} {85}},\ \bibinfo {pages} {949} (\bibinfo {year} {2012})}\BibitemShut
	{NoStop}%
	\bibitem [{\citenamefont {Goian}\ \emph {et~al.}(2009)\citenamefont {Goian},
		\citenamefont {Kamba}, \citenamefont {Hlinka}, \citenamefont {Van{\v{e}}k},
		\citenamefont {Belik}, \citenamefont {Kolodiazhnyi},\ and\ \citenamefont
		{Petzelt}}]{Goian09}%
	\BibitemOpen
	\bibfield  {author} {\bibinfo {author} {\bibfnamefont {V.}~\bibnamefont
			{Goian}}, \bibinfo {author} {\bibfnamefont {S.}~\bibnamefont {Kamba}},
		\bibinfo {author} {\bibfnamefont {J.}~\bibnamefont {Hlinka}}, \bibinfo
		{author} {\bibfnamefont {P.}~\bibnamefont {Van{\v{e}}k}}, \bibinfo {author}
		{\bibfnamefont {A.}~\bibnamefont {Belik}}, \bibinfo {author} {\bibfnamefont
			{T.}~\bibnamefont {Kolodiazhnyi}},\ and\ \bibinfo {author} {\bibfnamefont
			{J.}~\bibnamefont {Petzelt}},\ }\bibfield  {title} {\bibinfo {title} {Polar
			phonon mixing in magnetoelectric {EuTiO$_3$}},\ }\href@noop {} {\bibfield
		{journal} {\bibinfo  {journal} {Eur. Phys. J. B}\ }\textbf {\bibinfo {volume}
			{71}},\ \bibinfo {pages} {429} (\bibinfo {year} {2009})}\BibitemShut
	{NoStop}%
	\bibitem [{\citenamefont {Roleder}\ \emph {et~al.}(1996)\citenamefont
		{Roleder}, \citenamefont {Maglione}, \citenamefont {Fontana},\ and\
		\citenamefont {Dec}}]{Roleder96}%
	\BibitemOpen
	\bibfield  {author} {\bibinfo {author} {\bibfnamefont {K.}~\bibnamefont
			{Roleder}}, \bibinfo {author} {\bibfnamefont {M.}~\bibnamefont {Maglione}},
		\bibinfo {author} {\bibfnamefont {M.}~\bibnamefont {Fontana}},\ and\ \bibinfo
		{author} {\bibfnamefont {J.}~\bibnamefont {Dec}},\ }\bibfield  {title}
	{\bibinfo {title} {Behaviour of a polar relaxation mode around the phase
			transition point in the antiferroelectric single crystal},\ }\href@noop {}
	{\bibfield  {journal} {\bibinfo  {journal} {J. Condens. Matter Phys.}\
		}\textbf {\bibinfo {volume} {8}},\ \bibinfo {pages} {10669} (\bibinfo {year}
		{1996})}\BibitemShut {NoStop}%
	\bibitem [{\citenamefont {McGuire}\ \emph {et~al.}(1966)\citenamefont
		{McGuire}, \citenamefont {Shafer}, \citenamefont {Joenk}, \citenamefont
		{Alperin},\ and\ \citenamefont {Pickart}}]{McGuire66}%
	\BibitemOpen
	\bibfield  {author} {\bibinfo {author} {\bibfnamefont {T.}~\bibnamefont
			{McGuire}}, \bibinfo {author} {\bibfnamefont {M.}~\bibnamefont {Shafer}},
		\bibinfo {author} {\bibfnamefont {R.}~\bibnamefont {Joenk}}, \bibinfo
		{author} {\bibfnamefont {H.}~\bibnamefont {Alperin}},\ and\ \bibinfo {author}
		{\bibfnamefont {S.}~\bibnamefont {Pickart}},\ }\bibfield  {title} {\bibinfo
		{title} {Magnetic structure of {EuTiO$_3$}},\ }\href@noop {} {\bibfield
		{journal} {\bibinfo  {journal} {J. Appl. Phys.}\ }\textbf {\bibinfo {volume}
			{37}},\ \bibinfo {pages} {981} (\bibinfo {year} {1966})}\BibitemShut
	{NoStop}%
	\bibitem [{\citenamefont {Scagnoli}\ \emph {et~al.}(2012)\citenamefont
		{Scagnoli}, \citenamefont {Allieta}, \citenamefont {Walker}, \citenamefont
		{Scavini}, \citenamefont {Katsufuji}, \citenamefont {Sagarna}, \citenamefont
		{Zaharko},\ and\ \citenamefont {Mazzoli}}]{Scagnoli12}%
	\BibitemOpen
	\bibfield  {author} {\bibinfo {author} {\bibfnamefont {V.}~\bibnamefont
			{Scagnoli}}, \bibinfo {author} {\bibfnamefont {M.}~\bibnamefont {Allieta}},
		\bibinfo {author} {\bibfnamefont {H.}~\bibnamefont {Walker}}, \bibinfo
		{author} {\bibfnamefont {M.}~\bibnamefont {Scavini}}, \bibinfo {author}
		{\bibfnamefont {T.}~\bibnamefont {Katsufuji}}, \bibinfo {author}
		{\bibfnamefont {L.}~\bibnamefont {Sagarna}}, \bibinfo {author} {\bibfnamefont
			{O.}~\bibnamefont {Zaharko}},\ and\ \bibinfo {author} {\bibfnamefont
			{C.}~\bibnamefont {Mazzoli}},\ }\bibfield  {title} {\bibinfo {title}
		{{EuTiO$_3$} magnetic structure studied by neutron powder diffraction and
			resonant {X}-ray scattering},\ }\href@noop {} {\bibfield  {journal} {\bibinfo
			{journal} {Phys. Rev. B}\ }\textbf {\bibinfo {volume} {86}},\ \bibinfo
		{pages} {094432} (\bibinfo {year} {2012})}\BibitemShut {NoStop}%
	\bibitem [{\citenamefont {Ryan}\ \emph {et~al.}(2013)\citenamefont {Ryan},
		\citenamefont {Kim}, \citenamefont {Birol}, \citenamefont {Thompson},
		\citenamefont {Lee}, \citenamefont {Ke}, \citenamefont {Normile},
		\citenamefont {Karapetrova}, \citenamefont {Schiffer}, \citenamefont {Brown}
		\emph {et~al.}}]{Ryan13}%
	\BibitemOpen
	\bibfield  {author} {\bibinfo {author} {\bibfnamefont {P.}~\bibnamefont
			{Ryan}}, \bibinfo {author} {\bibfnamefont {J.}~\bibnamefont {Kim}}, \bibinfo
		{author} {\bibfnamefont {T.}~\bibnamefont {Birol}}, \bibinfo {author}
		{\bibfnamefont {P.}~\bibnamefont {Thompson}}, \bibinfo {author}
		{\bibfnamefont {J.}~\bibnamefont {Lee}}, \bibinfo {author} {\bibfnamefont
			{X.}~\bibnamefont {Ke}}, \bibinfo {author} {\bibfnamefont {P.}~\bibnamefont
			{Normile}}, \bibinfo {author} {\bibfnamefont {E.}~\bibnamefont
			{Karapetrova}}, \bibinfo {author} {\bibfnamefont {P.}~\bibnamefont
			{Schiffer}}, \bibinfo {author} {\bibfnamefont {S.}~\bibnamefont {Brown}},
		\emph {et~al.},\ }\bibfield  {title} {\bibinfo {title} {Reversible control of
			magnetic interactions by electric field in a single-phase material},\
	}\href@noop {} {\bibfield  {journal} {\bibinfo  {journal} {Nature Commun.}\
		}\textbf {\bibinfo {volume} {4}},\ \bibinfo {pages} {1334} (\bibinfo {year}
		{2013})}\BibitemShut {NoStop}%
	\bibitem [{\citenamefont {von Hippel}(1950)}]{Hippel50}%
	\BibitemOpen
	\bibfield  {author} {\bibinfo {author} {\bibfnamefont {A.}~\bibnamefont {von
				Hippel}},\ }\bibfield  {title} {\bibinfo {title} {Ferroelectricity, domain
			structure, and phase transitions of barium titanate},\ }\href@noop {}
	{\bibfield  {journal} {\bibinfo  {journal} {Rev. Mod. Phys.}\ }\textbf
		{\bibinfo {volume} {22}},\ \bibinfo {pages} {221} (\bibinfo {year}
		{1950})}\BibitemShut {NoStop}%
	\bibitem [{\citenamefont {Katsufuji}\ and\ \citenamefont
		{Takagi}(2001)}]{Katsufuji01}%
	\BibitemOpen
	\bibfield  {author} {\bibinfo {author} {\bibfnamefont {T.}~\bibnamefont
			{Katsufuji}}\ and\ \bibinfo {author} {\bibfnamefont {H.}~\bibnamefont
			{Takagi}},\ }\bibfield  {title} {\bibinfo {title} {Coupling between magnetism
			and dielectric properties in quantum paraelectric {EuTiO$_3$}},\ }\href@noop
	{} {\bibfield  {journal} {\bibinfo  {journal} {Phys. Rev. B}\ }\textbf
		{\bibinfo {volume} {64}},\ \bibinfo {pages} {054415} (\bibinfo {year}
		{2001})}\BibitemShut {NoStop}%
	\bibitem [{\citenamefont {Rushchanskii}\ \emph {et~al.}(2010)\citenamefont
		{Rushchanskii}, \citenamefont {Kamba}, \citenamefont {Goian}, \citenamefont
		{Van{\v{e}}k}, \citenamefont {Savinov}, \citenamefont {Prokle{\v{s}}ka},
		\citenamefont {Nuzhnyy}, \citenamefont {Kn{\'\i}{\v{z}}ek}, \citenamefont
		{Laufek}, \citenamefont {Eckel} \emph {et~al.}}]{Rushchanskii10}%
	\BibitemOpen
	\bibfield  {author} {\bibinfo {author} {\bibfnamefont {K.}~\bibnamefont
			{Rushchanskii}}, \bibinfo {author} {\bibfnamefont {S.}~\bibnamefont {Kamba}},
		\bibinfo {author} {\bibfnamefont {V.}~\bibnamefont {Goian}}, \bibinfo
		{author} {\bibfnamefont {P.}~\bibnamefont {Van{\v{e}}k}}, \bibinfo {author}
		{\bibfnamefont {M.}~\bibnamefont {Savinov}}, \bibinfo {author} {\bibfnamefont
			{J.}~\bibnamefont {Prokle{\v{s}}ka}}, \bibinfo {author} {\bibfnamefont
			{D.}~\bibnamefont {Nuzhnyy}}, \bibinfo {author} {\bibfnamefont
			{K.}~\bibnamefont {Kn{\'\i}{\v{z}}ek}}, \bibinfo {author} {\bibfnamefont
			{F.}~\bibnamefont {Laufek}}, \bibinfo {author} {\bibfnamefont
			{S.}~\bibnamefont {Eckel}}, \emph {et~al.},\ }\bibfield  {title} {\bibinfo
		{title} {A multiferroic material to search for the permanent electric dipole
			moment of the electron},\ }\href@noop {} {\bibfield  {journal} {\bibinfo
			{journal} {Nature Mater.}\ }\textbf {\bibinfo {volume} {9}},\ \bibinfo
		{pages} {649} (\bibinfo {year} {2010})}\BibitemShut {NoStop}%
	\bibitem [{\citenamefont {Goian}\ \emph {et~al.}(2013)\citenamefont {Goian},
		\citenamefont {Kamba}, \citenamefont {Van{\v{e}}k}, \citenamefont {Savinov},
		\citenamefont {Kadlec},\ and\ \citenamefont {Prokle{\v{s}}ka}}]{Goian13}%
	\BibitemOpen
	\bibfield  {author} {\bibinfo {author} {\bibfnamefont {V.}~\bibnamefont
			{Goian}}, \bibinfo {author} {\bibfnamefont {S.}~\bibnamefont {Kamba}},
		\bibinfo {author} {\bibfnamefont {P.}~\bibnamefont {Van{\v{e}}k}}, \bibinfo
		{author} {\bibfnamefont {M.}~\bibnamefont {Savinov}}, \bibinfo {author}
		{\bibfnamefont {C.}~\bibnamefont {Kadlec}},\ and\ \bibinfo {author}
		{\bibfnamefont {J.}~\bibnamefont {Prokle{\v{s}}ka}},\ }\bibfield  {title}
	{\bibinfo {title} {Magnetic and dielectric properties of multiferroic
			{Eu$_{0.5}$Ba$_{0.25}$Sr$_{0.25}$TiO$_3$} ceramics},\ }\href@noop {}
	{\bibfield  {journal} {\bibinfo  {journal} {Phase Transit.}\ }\textbf
		{\bibinfo {volume} {86}},\ \bibinfo {pages} {191} (\bibinfo {year}
		{2013})}\BibitemShut {NoStop}%
	\bibitem [{\citenamefont {Bussmann-Holder}\ and\ \citenamefont
		{K{\"o}hler}(2015)}]{Bussmann-Holder15}%
	\BibitemOpen
	\bibfield  {author} {\bibinfo {author} {\bibfnamefont {A.}~\bibnamefont
			{Bussmann-Holder}}\ and\ \bibinfo {author} {\bibfnamefont {J.}~\bibnamefont
			{K{\"o}hler}},\ }\bibfield  {title} {\bibinfo {title} {Revisiting the
			fascinating properties of {EuTiO$_3$} and its mixed crystals with
			{SrTiO$_3$}: Possible candidates for novel functionalities},\ }\href@noop {}
	{\bibfield  {journal} {\bibinfo  {journal} {J. Phys. Chem. Solids}\ }\textbf
		{\bibinfo {volume} {84}},\ \bibinfo {pages} {2} (\bibinfo {year}
		{2015})}\BibitemShut {NoStop}%
	\bibitem [{Sup()}]{Suppl}%
	\BibitemOpen
	\href@noop {} {\bibinfo {title} {See {Supplemental} material at [{URL} will
			be inserted by publisher] for figures of sample preparation, temperature
			dependence of the lattice parameter, further heat capacity measurements, and
			additional dielectric and magnetic properties of the {EBSTO} ceramics and
			single crystal.}}\BibitemShut {Stop}%
	\bibitem [{\citenamefont {{\v{S}}milauerov{\'a}}\ \emph
		{et~al.}(2014)\citenamefont {{\v{S}}milauerov{\'a}}, \citenamefont
		{Posp{\'\i}{\v{s}}il}, \citenamefont {Harcuba}, \citenamefont {Hol{\'y}},\
		and\ \citenamefont {Jane{\v{c}}ek}}]{Smilauerova14}%
	\BibitemOpen
	\bibfield  {author} {\bibinfo {author} {\bibfnamefont {J.}~\bibnamefont
			{{\v{S}}milauerov{\'a}}}, \bibinfo {author} {\bibfnamefont {J.}~\bibnamefont
			{Posp{\'\i}{\v{s}}il}}, \bibinfo {author} {\bibfnamefont {P.}~\bibnamefont
			{Harcuba}}, \bibinfo {author} {\bibfnamefont {V.}~\bibnamefont {Hol{\'y}}},\
		and\ \bibinfo {author} {\bibfnamefont {M.}~\bibnamefont {Jane{\v{c}}ek}},\
	}\bibfield  {title} {\bibinfo {title} {Single crystal growth of {TIMETAL}
			{LCB} titanium alloy by a floating zone method},\ }\href@noop {} {\bibfield
		{journal} {\bibinfo  {journal} {J. Cryst. Growth}\ }\textbf {\bibinfo
			{volume} {405}},\ \bibinfo {pages} {92} (\bibinfo {year} {2014})}\BibitemShut
	{NoStop}%
	\bibitem [{\citenamefont {Zajic}\ \emph {et~al.}(2023)\citenamefont {Zajic},
		\citenamefont {Klejch}, \citenamefont {Elias}, \citenamefont {Klicpera},
		\citenamefont {Beitlerov{\'a}}, \citenamefont {Nikl},\ and\ \citenamefont
		{Pospisil}}]{Zajic23}%
	\BibitemOpen
	\bibfield  {author} {\bibinfo {author} {\bibfnamefont {F.}~\bibnamefont
			{Zajic}}, \bibinfo {author} {\bibfnamefont {M.}~\bibnamefont {Klejch}},
		\bibinfo {author} {\bibfnamefont {A.}~\bibnamefont {Elias}}, \bibinfo
		{author} {\bibfnamefont {M.}~\bibnamefont {Klicpera}}, \bibinfo {author}
		{\bibfnamefont {A.}~\bibnamefont {Beitlerov{\'a}}}, \bibinfo {author}
		{\bibfnamefont {M.}~\bibnamefont {Nikl}},\ and\ \bibinfo {author}
		{\bibfnamefont {J.}~\bibnamefont {Pospisil}},\ }\bibfield  {title} {\bibinfo
		{title} {{Nd}:{YAG} single crystals grown by the floating zone method in a
			laser furnace},\ }\href@noop {} {\bibfield  {journal} {\bibinfo  {journal}
			{Cryst. Growth Des.}\ }\textbf {\bibinfo {volume} {23}},\ \bibinfo {pages}
		{2609} (\bibinfo {year} {2023})}\BibitemShut {NoStop}%
	\bibitem [{\citenamefont {Kriegner}\ \emph {et~al.}(2015)\citenamefont
		{Kriegner}, \citenamefont {Mat{\v{e}}j}, \citenamefont {Ku{\v{z}}el},\ and\
		\citenamefont {Hol{\'y}}}]{Kriegner15}%
	\BibitemOpen
	\bibfield  {author} {\bibinfo {author} {\bibfnamefont {D.}~\bibnamefont
			{Kriegner}}, \bibinfo {author} {\bibfnamefont {Z.}~\bibnamefont
			{Mat{\v{e}}j}}, \bibinfo {author} {\bibfnamefont {R.}~\bibnamefont
			{Ku{\v{z}}el}},\ and\ \bibinfo {author} {\bibfnamefont {V.}~\bibnamefont
			{Hol{\'y}}},\ }\bibfield  {title} {\bibinfo {title} {Powder diffraction in
			{B}ragg--{B}rentano geometry with straight linear detectors},\ }\href@noop {}
	{\bibfield  {journal} {\bibinfo  {journal} {J. Appl. Crystallogr.}\ }\textbf
		{\bibinfo {volume} {48}},\ \bibinfo {pages} {613} (\bibinfo {year}
		{2015})}\BibitemShut {NoStop}%
	\bibitem [{\citenamefont {Fujiwara}\ \emph {et~al.}(2007)\citenamefont
		{Fujiwara}, \citenamefont {Matsumoto}, \citenamefont {Nakazawab},
		\citenamefont {Hisada},\ and\ \citenamefont {Uwatoko}}]{Fujiwara07}%
	\BibitemOpen
	\bibfield  {author} {\bibinfo {author} {\bibfnamefont {N.}~\bibnamefont
			{Fujiwara}}, \bibinfo {author} {\bibfnamefont {T.}~\bibnamefont {Matsumoto}},
		\bibinfo {author} {\bibfnamefont {K.}~\bibnamefont {Nakazawab}}, \bibinfo
		{author} {\bibfnamefont {A.}~\bibnamefont {Hisada}},\ and\ \bibinfo {author}
		{\bibfnamefont {Y.}~\bibnamefont {Uwatoko}},\ }\bibfield  {title} {\bibinfo
		{title} {Fabrication and efficiency evaluation of a hybrid {NiCrAl} pressure
			cell up to 4 {GPa}},\ }\href@noop {} {\bibfield  {journal} {\bibinfo
			{journal} {Rev. Sci. Instrum.}\ }\textbf {\bibinfo {volume} {78}},\ \bibinfo
		{pages} {73905} (\bibinfo {year} {2007})}\BibitemShut {NoStop}%
	\bibitem [{\citenamefont {Yokogawa}\ \emph {et~al.}(2007)\citenamefont
		{Yokogawa}, \citenamefont {Murata}, \citenamefont {Yoshino},\ and\
		\citenamefont {Aoyama}}]{Yokogawa07}%
	\BibitemOpen
	\bibfield  {author} {\bibinfo {author} {\bibfnamefont {K.}~\bibnamefont
			{Yokogawa}}, \bibinfo {author} {\bibfnamefont {K.}~\bibnamefont {Murata}},
		\bibinfo {author} {\bibfnamefont {H.}~\bibnamefont {Yoshino}},\ and\ \bibinfo
		{author} {\bibfnamefont {S.}~\bibnamefont {Aoyama}},\ }\bibfield  {title}
	{\bibinfo {title} {Solidification of high-pressure medium {D}aphne 7373},\
	}\href@noop {} {\bibfield  {journal} {\bibinfo  {journal} {Jpn. J. Appl.
				Phys.}\ }\textbf {\bibinfo {volume} {46}},\ \bibinfo {pages} {3636} (\bibinfo
		{year} {2007})}\BibitemShut {NoStop}%
	\bibitem [{\citenamefont {Murata}\ \emph {et~al.}(1997)\citenamefont {Murata},
		\citenamefont {Yoshino}, \citenamefont {Yadav}, \citenamefont {Honda},\ and\
		\citenamefont {Shirakawa}}]{Murata97}%
	\BibitemOpen
	\bibfield  {author} {\bibinfo {author} {\bibfnamefont {K.}~\bibnamefont
			{Murata}}, \bibinfo {author} {\bibfnamefont {H.}~\bibnamefont {Yoshino}},
		\bibinfo {author} {\bibfnamefont {H.~O.}\ \bibnamefont {Yadav}}, \bibinfo
		{author} {\bibfnamefont {Y.}~\bibnamefont {Honda}},\ and\ \bibinfo {author}
		{\bibfnamefont {N.}~\bibnamefont {Shirakawa}},\ }\bibfield  {title} {\bibinfo
		{title} {{Pt} resistor thermometry and pressure calibration in a clamped
			pressure cell with the medium, {D}aphne 7373},\ }\href@noop {} {\bibfield
		{journal} {\bibinfo  {journal} {Rev. Sci. Instrum.}\ }\textbf {\bibinfo
			{volume} {68}},\ \bibinfo {pages} {2490} (\bibinfo {year}
		{1997})}\BibitemShut {NoStop}%
	\bibitem [{\citenamefont {Sta{\v{s}}ko}\ \emph {et~al.}(2020)\citenamefont
		{Sta{\v{s}}ko}, \citenamefont {Prchal}, \citenamefont {Klicpera},
		\citenamefont {Aoki},\ and\ \citenamefont {Murata}}]{Stasko20}%
	\BibitemOpen
	\bibfield  {author} {\bibinfo {author} {\bibfnamefont {D.}~\bibnamefont
			{Sta{\v{s}}ko}}, \bibinfo {author} {\bibfnamefont {J.}~\bibnamefont
			{Prchal}}, \bibinfo {author} {\bibfnamefont {M.}~\bibnamefont {Klicpera}},
		\bibinfo {author} {\bibfnamefont {S.}~\bibnamefont {Aoki}},\ and\ \bibinfo
		{author} {\bibfnamefont {K.}~\bibnamefont {Murata}},\ }\bibfield  {title}
	{\bibinfo {title} {Pressure media for high pressure experiments, {D}aphne oil
			7000 series},\ }\href@noop {} {\bibfield  {journal} {\bibinfo  {journal}
			{High Press. Res.}\ }\textbf {\bibinfo {volume} {40}},\ \bibinfo {pages}
		{525} (\bibinfo {year} {2020})}\BibitemShut {NoStop}%
	\bibitem [{\citenamefont {Barrett}(1952)}]{Barrett52}%
	\BibitemOpen
	\bibfield  {author} {\bibinfo {author} {\bibfnamefont {J.~H.}\ \bibnamefont
			{Barrett}},\ }\bibfield  {title} {\bibinfo {title} {Dielectric constant in
			perovskite type crystals},\ }\href@noop {} {\bibfield  {journal} {\bibinfo
			{journal} {Phys. Rev.}\ }\textbf {\bibinfo {volume} {86}},\ \bibinfo {pages}
		{118} (\bibinfo {year} {1952})}\BibitemShut {NoStop}%
	\bibitem [{\citenamefont {Rep{\v{c}}ek}\ \emph {et~al.}(2020)\citenamefont
		{Rep{\v{c}}ek}, \citenamefont {Kadlec}, \citenamefont {Kadlec}, \citenamefont
		{Savinov}, \citenamefont {Kachl{\'\i}k}, \citenamefont {Drahokoupil},
		\citenamefont {Proschek}, \citenamefont {Prokle{\v{s}}ka}, \citenamefont
		{Maca},\ and\ \citenamefont {Kamba}}]{Repcek20}%
	\BibitemOpen
	\bibfield  {author} {\bibinfo {author} {\bibfnamefont {D.}~\bibnamefont
			{Rep{\v{c}}ek}}, \bibinfo {author} {\bibfnamefont {C.}~\bibnamefont
			{Kadlec}}, \bibinfo {author} {\bibfnamefont {F.}~\bibnamefont {Kadlec}},
		\bibinfo {author} {\bibfnamefont {M.}~\bibnamefont {Savinov}}, \bibinfo
		{author} {\bibfnamefont {M.}~\bibnamefont {Kachl{\'\i}k}}, \bibinfo {author}
		{\bibfnamefont {J.}~\bibnamefont {Drahokoupil}}, \bibinfo {author}
		{\bibfnamefont {P.}~\bibnamefont {Proschek}}, \bibinfo {author}
		{\bibfnamefont {J.}~\bibnamefont {Prokle{\v{s}}ka}}, \bibinfo {author}
		{\bibfnamefont {K.}~\bibnamefont {Maca}},\ and\ \bibinfo {author}
		{\bibfnamefont {S.}~\bibnamefont {Kamba}},\ }\bibfield  {title} {\bibinfo
		{title} {Seemingly anisotropic magnetodielectric effect in isotropic
			{EuTiO$_3$} ceramics},\ }\href@noop {} {\bibfield  {journal} {\bibinfo
			{journal} {Phys. Rev. B}\ }\textbf {\bibinfo {volume} {102}},\ \bibinfo
		{pages} {144402} (\bibinfo {year} {2020})}\BibitemShut {NoStop}%
	\bibitem [{\citenamefont {Fischer}\ and\ \citenamefont
		{Hegenbarth}(1985)}]{Fischer85}%
	\BibitemOpen
	\bibfield  {author} {\bibinfo {author} {\bibfnamefont {E.}~\bibnamefont
			{Fischer}}\ and\ \bibinfo {author} {\bibfnamefont {E.}~\bibnamefont
			{Hegenbarth}},\ }\bibfield  {title} {\bibinfo {title} {Glasslike behaviour of
			{SrTiO$_3$} single crystal at low temperatures},\ }\href@noop {} {\bibfield
		{journal} {\bibinfo  {journal} {Ferroel. Lett. Sect.}\ }\textbf {\bibinfo
			{volume} {5}},\ \bibinfo {pages} {21} (\bibinfo {year} {1985})}\BibitemShut
	{NoStop}%
	\bibitem [{\citenamefont {Vendik}\ and\ \citenamefont
		{Zubko}(1997)}]{Vendik97}%
	\BibitemOpen
	\bibfield  {author} {\bibinfo {author} {\bibfnamefont {O.~G.}\ \bibnamefont
			{Vendik}}\ and\ \bibinfo {author} {\bibfnamefont {S.~P.}\ \bibnamefont
			{Zubko}},\ }\bibfield  {title} {\bibinfo {title} {Modeling the dielectric
			response of incipient ferroelectrics},\ }\href@noop {} {\bibfield  {journal}
		{\bibinfo  {journal} {J. Appl. Phys.}\ }\textbf {\bibinfo {volume} {82}},\
		\bibinfo {pages} {4475} (\bibinfo {year} {1997})}\BibitemShut {NoStop}%
	\bibitem [{\citenamefont {Coak}\ \emph {et~al.}(2019)\citenamefont {Coak},
		\citenamefont {Haines}, \citenamefont {Liu}, \citenamefont
		{Guzm{\'a}n-Verri},\ and\ \citenamefont {Saxena}}]{Coak19}%
	\BibitemOpen
	\bibfield  {author} {\bibinfo {author} {\bibfnamefont {M.~J.}\ \bibnamefont
			{Coak}}, \bibinfo {author} {\bibfnamefont {C.~R.~S.}\ \bibnamefont {Haines}},
		\bibinfo {author} {\bibfnamefont {C.}~\bibnamefont {Liu}}, \bibinfo {author}
		{\bibfnamefont {G.~G.}\ \bibnamefont {Guzm{\'a}n-Verri}},\ and\ \bibinfo
		{author} {\bibfnamefont {S.~S.}\ \bibnamefont {Saxena}},\ }\bibfield  {title}
	{\bibinfo {title} {Pressure dependence of ferroelectric quantum critical
			fluctuations},\ }\href@noop {} {\bibfield  {journal} {\bibinfo  {journal}
			{Phys. Rev. B}\ }\textbf {\bibinfo {volume} {100}},\ \bibinfo {pages}
		{214111} (\bibinfo {year} {2019})}\BibitemShut {NoStop}%
	\bibitem [{\citenamefont {Coak}\ \emph {et~al.}(2020)\citenamefont {Coak},
		\citenamefont {Haines}, \citenamefont {Liu}, \citenamefont {Rowley},
		\citenamefont {Lonzarich},\ and\ \citenamefont {Saxena}}]{Coak20}%
	\BibitemOpen
	\bibfield  {author} {\bibinfo {author} {\bibfnamefont {M.~J.}\ \bibnamefont
			{Coak}}, \bibinfo {author} {\bibfnamefont {C.~R.}\ \bibnamefont {Haines}},
		\bibinfo {author} {\bibfnamefont {C.}~\bibnamefont {Liu}}, \bibinfo {author}
		{\bibfnamefont {S.~E.}\ \bibnamefont {Rowley}}, \bibinfo {author}
		{\bibfnamefont {G.~G.}\ \bibnamefont {Lonzarich}},\ and\ \bibinfo {author}
		{\bibfnamefont {S.~S.}\ \bibnamefont {Saxena}},\ }\bibfield  {title}
	{\bibinfo {title} {Quantum critical phenomena in a compressible displacive
			ferroelectric},\ }\href@noop {} {\bibfield  {journal} {\bibinfo  {journal}
			{PNAS}\ }\textbf {\bibinfo {volume} {117}},\ \bibinfo {pages} {12707}
		(\bibinfo {year} {2020})}\BibitemShut {NoStop}%
	\bibitem [{\citenamefont {Petrovi{\'c}}\ \emph {et~al.}(2013)\citenamefont
		{Petrovi{\'c}}, \citenamefont {Kato}, \citenamefont {Sunku}, \citenamefont
		{Ito}, \citenamefont {Sengupta}, \citenamefont {Spalek}, \citenamefont
		{Shimuta}, \citenamefont {Katsufuji}, \citenamefont {Batista}, \citenamefont
		{Saxena} \emph {et~al.}}]{Petrovic13}%
	\BibitemOpen
	\bibfield  {author} {\bibinfo {author} {\bibfnamefont {A.}~\bibnamefont
			{Petrovi{\'c}}}, \bibinfo {author} {\bibfnamefont {Y.}~\bibnamefont {Kato}},
		\bibinfo {author} {\bibfnamefont {S.}~\bibnamefont {Sunku}}, \bibinfo
		{author} {\bibfnamefont {T.}~\bibnamefont {Ito}}, \bibinfo {author}
		{\bibfnamefont {P.}~\bibnamefont {Sengupta}}, \bibinfo {author}
		{\bibfnamefont {L.}~\bibnamefont {Spalek}}, \bibinfo {author} {\bibfnamefont
			{M.}~\bibnamefont {Shimuta}}, \bibinfo {author} {\bibfnamefont
			{T.}~\bibnamefont {Katsufuji}}, \bibinfo {author} {\bibfnamefont
			{C.}~\bibnamefont {Batista}}, \bibinfo {author} {\bibfnamefont
			{S.}~\bibnamefont {Saxena}}, \emph {et~al.},\ }\bibfield  {title} {\bibinfo
		{title} {Electric field modulation of the tetragonal domain orientation
			revealed in the magnetic ground state of quantum paraelectric {EuTiO$_3$}},\
	}\href@noop {} {\bibfield  {journal} {\bibinfo  {journal} {Phys. Rev. B}\
		}\textbf {\bibinfo {volume} {87}},\ \bibinfo {pages} {064103} (\bibinfo
		{year} {2013})}\BibitemShut {NoStop}%
	\bibitem [{exp()}]{explanation}%
	\BibitemOpen
	\href@noop {} {\bibinfo {title} {$\chi_{\boldsymbol{\uppsi}}({T})$ data
			measured using {Hall} probes are noisier than those obtained from {VSM}.
			{C}onsequently, $\chi_{\boldsymbol{\uppsi}}^{-1} ({T})$ from {H}all probes is
			not representative enough above 4 {K}. {T}herefore we present joint data of
			{VSM} and {H}all probe techniques in the inset of {F}ig.
			\ref{fig:magn_susc_and_inv_magn_susc_on_T}}}\BibitemShut {NoStop}%
	\bibitem [{\citenamefont {Jaoui}\ \emph {et~al.}(2023)\citenamefont {Jaoui},
		\citenamefont {Jiang}, \citenamefont {Li}, \citenamefont {Tomioka},
		\citenamefont {Inoue}, \citenamefont {Engelmayer}, \citenamefont {Sharma},
		\citenamefont {P{\"a}tzold}, \citenamefont {Lorenz}, \citenamefont
		{Fauqu{\'e}} \emph {et~al.}}]{Jaoui23}%
	\BibitemOpen
	\bibfield  {author} {\bibinfo {author} {\bibfnamefont {A.}~\bibnamefont
			{Jaoui}}, \bibinfo {author} {\bibfnamefont {S.}~\bibnamefont {Jiang}},
		\bibinfo {author} {\bibfnamefont {X.}~\bibnamefont {Li}}, \bibinfo {author}
		{\bibfnamefont {Y.}~\bibnamefont {Tomioka}}, \bibinfo {author} {\bibfnamefont
			{I.~H.}\ \bibnamefont {Inoue}}, \bibinfo {author} {\bibfnamefont
			{J.}~\bibnamefont {Engelmayer}}, \bibinfo {author} {\bibfnamefont
			{R.}~\bibnamefont {Sharma}}, \bibinfo {author} {\bibfnamefont
			{L.}~\bibnamefont {P{\"a}tzold}}, \bibinfo {author} {\bibfnamefont
			{T.}~\bibnamefont {Lorenz}}, \bibinfo {author} {\bibfnamefont
			{B.}~\bibnamefont {Fauqu{\'e}}}, \emph {et~al.},\ }\bibfield  {title}
	{\bibinfo {title} {Glasslike thermal conductivity and narrow insulating gap
			of {EuTiO$_3$}},\ }\href@noop {} {\bibfield  {journal} {\bibinfo  {journal}
			{Phys. Rev. Mat.}\ }\textbf {\bibinfo {volume} {7}},\ \bibinfo {pages}
		{094604} (\bibinfo {year} {2023})}\BibitemShut {NoStop}%
	\bibitem [{\citenamefont {Beauvillain}\ \emph {et~al.}(1978)\citenamefont
		{Beauvillain}, \citenamefont {Renard}, \citenamefont {Laursen},\ and\
		\citenamefont {Walker}}]{Beauvillain78}%
	\BibitemOpen
	\bibfield  {author} {\bibinfo {author} {\bibfnamefont {P.}~\bibnamefont
			{Beauvillain}}, \bibinfo {author} {\bibfnamefont {J.-P.}\ \bibnamefont
			{Renard}}, \bibinfo {author} {\bibfnamefont {I.}~\bibnamefont {Laursen}},\
		and\ \bibinfo {author} {\bibfnamefont {P.}~\bibnamefont {Walker}},\
	}\bibfield  {title} {\bibinfo {title} {Critical behavior of the magnetic
			susceptibility of the uniaxial ferromagnet {LiHoF$_4$}},\ }\href@noop {}
	{\bibfield  {journal} {\bibinfo  {journal} {Phys. Rev. B}\ }\textbf {\bibinfo
			{volume} {18}},\ \bibinfo {pages} {3360} (\bibinfo {year}
		{1978})}\BibitemShut {NoStop}%
	\bibitem [{\citenamefont {Bitko}\ \emph {et~al.}(1996)\citenamefont {Bitko},
		\citenamefont {Rosenbaum},\ and\ \citenamefont {Aeppli}}]{Bitko96}%
	\BibitemOpen
	\bibfield  {author} {\bibinfo {author} {\bibfnamefont {D.}~\bibnamefont
			{Bitko}}, \bibinfo {author} {\bibfnamefont {T.}~\bibnamefont {Rosenbaum}},\
		and\ \bibinfo {author} {\bibfnamefont {G.}~\bibnamefont {Aeppli}},\
	}\bibfield  {title} {\bibinfo {title} {Quantum critical behavior for a model
			magnet},\ }\href@noop {} {\bibfield  {journal} {\bibinfo  {journal} {Phys.
				Rev. Lett.}\ }\textbf {\bibinfo {volume} {77}},\ \bibinfo {pages} {940}
		(\bibinfo {year} {1996})}\BibitemShut {NoStop}%
	\bibitem [{\citenamefont {Griffin}\ \emph {et~al.}(1980)\citenamefont
		{Griffin}, \citenamefont {Huster},\ and\ \citenamefont
		{Folweiler}}]{Griffin80}%
	\BibitemOpen
	\bibfield  {author} {\bibinfo {author} {\bibfnamefont {J.}~\bibnamefont
			{Griffin}}, \bibinfo {author} {\bibfnamefont {M.}~\bibnamefont {Huster}},\
		and\ \bibinfo {author} {\bibfnamefont {R.~J.}\ \bibnamefont {Folweiler}},\
	}\bibfield  {title} {\bibinfo {title} {Critical behavior of the spontaneous
			magnetization at marginal dimensionality in {LiHoF$_4$}},\ }\href@noop {}
	{\bibfield  {journal} {\bibinfo  {journal} {Phys. Rev. B}\ }\textbf {\bibinfo
			{volume} {22}},\ \bibinfo {pages} {4370} (\bibinfo {year}
		{1980})}\BibitemShut {NoStop}%
\end{thebibliography}
\end{document}